\documentclass[
aps,%
10pt,%
final,%
notitlepage,%
oneside,%
twocolumn,%
nobibnotes,%
nofootinbib,%
superscriptaddress,%
showpacs,%
centertags,%
showkeys,%
amssymb,%
]{revtex4}

  \usepackage[utf8]{inputenc}
  \usepackage{microtype} 
  \usepackage{graphicx}
    \graphicspath{{/}{figures/}}
	\DeclareGraphicsExtensions{.pdf,.png}
    \setkeys{Gin}{width=\linewidth,totalheight=\textheight,keepaspectratio} 
  \usepackage[caption=false]{subfig}
  \usepackage{xcolor}
  \usepackage{comment}
  \usepackage{amsmath,amssymb}
  \usepackage{tabularx,booktabs}
  \usepackage{todonotes}
  \usepackage{xspace}
  \usepackage{lipsum}
  \usepackage{units}
  \usepackage{bm}

\newcommand*\doctitle{Scientific paper}
\newcommand*\docauthor{Erik Bartoš}

\usepackage[pdftex]{hyperref}
\hypersetup{
      breaklinks    = true,
      baseurl       = http://,
      pdfborder     = 0 0 0,
      pdfpagemode   = UseNone,
      pdfstartpage  = 1,
      pdfcreator    = Pdf\LaTeX{} with hyperref package,
      pdfproducer   = Pdf\LaTeX{},
      bookmarksopen = true,
      pdfauthor     = {\docauthor},
      pdftitle      = {\doctitle},
      pdfsubject    = {Scientific paper},
      pdfkeywords   = {OANDA, string, model, time series, forecasting},
      unicode       = true,
      colorlinks    = true
}

\newcommand{\m}[1]{\mathrm {#1}}
\newcommand{\nn}{\nonumber}
\newcommand{\dd}{\m{d}}
\newcommand*{\bra}{\mathinner\langle}
\newcommand*{\ket}{\mathinner\rangle}
\newcommand*{\ack}{\mathinner\vert}

\newcounter{mycomment}

\newcounter{twocomment}

\makeatletter
\renewcommand\@dotsep{10000}
\makeatother

\begin{document}


\title{Identification of market trends with string and D2-brane maps}

\author{Erik Bartoš}%
\email{erik.bartos@savba.sk}
\affiliation{%
Institute of Physics, Slovak Academy of Sciences, Dúbravská cesta 9, 845 11 Bratislava, Slovak Republic
}%
\author{Richard Pinčák}
\email{pincak@saske.sk}
\affiliation{Institute of Experimental Physics, Slovak Academy of Sciences,\\ Watsonova 47, 043 53 Košice, Slovak Republic}
\affiliation{Bogoliubov Laboratory of Theoretical Physics,\\ Joint Institute for Nuclear Research, 141980 Dubna, Moscow Region, Russia}

\date{\today}

\begin{abstract}
The multi dimensional string objects are introduced as a new alternative for an application of string models for time series forecasting in trading on financial markets. The objects are represented by open string with 2-endpoints and D2-brane, which are continuous enhancement of 1-endpoint open string model. We show how new object properties can change the statistics of the predictors, which makes them the candidates for modeling a wide range of time series systems. String angular momentum is proposed as another tool to analyze the stability of currency rates except the historical volatility. To show the reliability of our approach with application of string models for time series forecasting we present the results of real demo simulations for four currency exchange pairs. 
\end{abstract}

\maketitle

\section{Introduction}\label{sec:intro}

Trading and predicting foreign exchange in the forex market \cite{Fang2003369} has become one of the intriguing topic and is extensively studied by researchers from different fields due to its commercial applications and attractive benefits that it has to offer. Algorithmic trading with large amount of processed data, the ticks in millisecond scale, requires new physical methods to describe the statistics of the return intervals on the short and large scales \cite{Lux:2000} and new geometric representation of data , e.~g., new view on data statistics in higher dimensions \cite{Kabin:2016}. Moreover, the global markets consist of a large number of interacting units and their time-averaged dynamics resemble the systems with many-body effects.
The classical statistical instruments which treats the market as a whole, like the returns and volatility distributions \cite{Wang:2009,Wang:2009b}, must be enhanced by new phenomena from informational and social sciences. Theoretical interest is also oriented to the distribution of the occurrence of rare extreme events in historical time series data \cite{Bunde:2005,Hartmann:2004}. Their clustering in data records indicates the existence of a long-term memory dependencies in financial time series, which is intensively studied \cite{Bogachev:2008}, i.~e., by multiplicative random cascade models. New approaches covering the findings of the long memory effects of forex data \cite{Nacher:2012} and its stochastic features in the presence of nonstationarity \cite{Anvari:2013}, the renormalization group approach \cite{Zamparo:2013}, exploitation of genetic algorithms \cite{Venkatesan2002625}, open novel perspectives.

We work on the concept which approach the string theory \cite{Zwiebach:2009} to the field of time series forecast and data analysis through a transformation of currency rate data to the topology of physical strings and branes \cite{Horvath:2012zz,Pincak:2013aa,Pincak:2014ba}. The ideas have been practically demonstrated by a novel prediction method based on string invariants \cite{Bundzel:2015bc} with genetic algorithm for optimization of method's parameters. The method has been tested on competition and real world data, its performance compared to artificial neural networks and support vector machines algorithms. Another interesting application has been the construction of trading algorithm based on 1-endpoint strings and the demonstration of model properties on real online trade system \cite{Pincak:2015hha}. Stability of the algorithm on transaction costs for long trade periods has been confirmed and compared to benchmark prediction models and trading strategies.

The aim of this paper is to outline new possibilities how the prediction models in trading on financial market can be enhanced in the framework of a string theory. We propose to proceed from simple 1-endpoint and 2-endpoints strings to more complex objects, D2-branes. The D2-branes have the ability to smooth the movement of prices on the market and to process the preserved market memory with better efficiency than in the case of the strings, the study of a statistics of momenta of string objects reveal the perspectives of D2-branes. However, the simulations with prediction models based on string approach show that one can profit only on the regions with high stability. In real data, the fluctuations of forex market prices brake the statistics of the predictions and one must build into the models various trading brakes, to deal with the rapid changes. The evaluation of a volatility \cite{Wang:2006,Gallo2015620} serves as one of the sources of analyzing tools in pricing strategies. We introduce new methodics based on the analogy with the angular momentum in the string theory. Its application into trading models can serve as complementary financial instrument in addition to a volatility. The changes of Regge slope parameter or the string tension can identify trends on the market, their understanding allows us to dynamically change an intra-string characterization (reduction a string length for a short period) and better predict the movement of prices. Especially in large market fluctuations, their exploitation needs further experimental verification.

The rest of the paper is organized as follows. Section~\ref{sec:strings} formulates the general models of multi dimensional string models. Their properties and comparison with previous models are discussed. In Section~\ref{sec:regge}, we introduce the Regge alpha slope for the investigation of the stability of currency rates. The obtained results are summarized in the last section. In Appendix~\ref{app:B} we demonstrate the application of our model for the real demo sessions for currency pairs EUR/USD, CHF/JPY, AUD/CAD, AUD/JPY.

\section{From simple to complex strings}\label{sec:strings}

The concept of string maps is based on the connection of the currency quotes and the string objects.
For the defined time series of currency exchange rates for the ask $p_{\m{ask}}(\tau)$ and bid $p_{\m{bid}}(\tau)$ values in time $\tau$ one can construct the string maps with the typical length $l_s$.  These non-local objects serve as the basic objects for further operations. In contrast to classical time series forecasting methods, e.~g., autoregressive and moving average models, which forecast the variable of interest using a linear combination of past values or errors of the variable, the string maps carry the  larger price history, thereafter the trends of irregular or untypical price changes can be caught with better accuracy.

In the works \cite{Horvath:2012zz,Pincak:2013aa} the $q$-deformed prediction model based on the deviations from benchmark string sequence of 1-endpoint string map $P_{q}^{(1)}(\tau,h)$ was thoroughly studied. The momentum $M$ of the string (the predictor) were proposed for the study of deviations of string maps from benchmark string sequence in the form
\begin{multline} \label{eq:mp}
M(l_s, m, q, \varphi) =\\ \left(\frac{1}{l_s+1}
\sum_{h=0}^{l_s}\Big| P_{N}(\tau,h) - F_{\m{CS}}(h,\varphi)\Big|^q\right)^{1/q},
\end{multline}
for $m,\;q>0$ , $l_s$ is the string length. $P_{N}(\tau,h)$ represents the generalized $N$-points string map, in previous case $P_{N}(\tau,h) \to P_{q}^{(1)}(\tau,h)$.  The regular function $F_{\m{CS}}(h,\varphi)$ could be substituted by various periodic functions. In \cite{Pincak:2015hha} we have used the form
\begin{equation} \label{eq:rf}
F_{\m{CS}}(h,\varphi) = \frac{1}{2}\big(1+\cos(\tilde{\varphi})\big),\quad \tilde{\varphi} = \frac{2\pi m h}{l_s + 1} + \varphi.
\end{equation}
We have shown that the final form of Eq.~(\ref{eq:rf}) depends on the trading strategy, when one looks the regions in the time series market with almost invariant values of $M$ and tries to predict the increase or decrease of prices with the best efficiency.

\begin{figure}[!tb]
	\centering
	\subfloat[1-endpoint open string]{
		\includegraphics[width=\columnwidth]{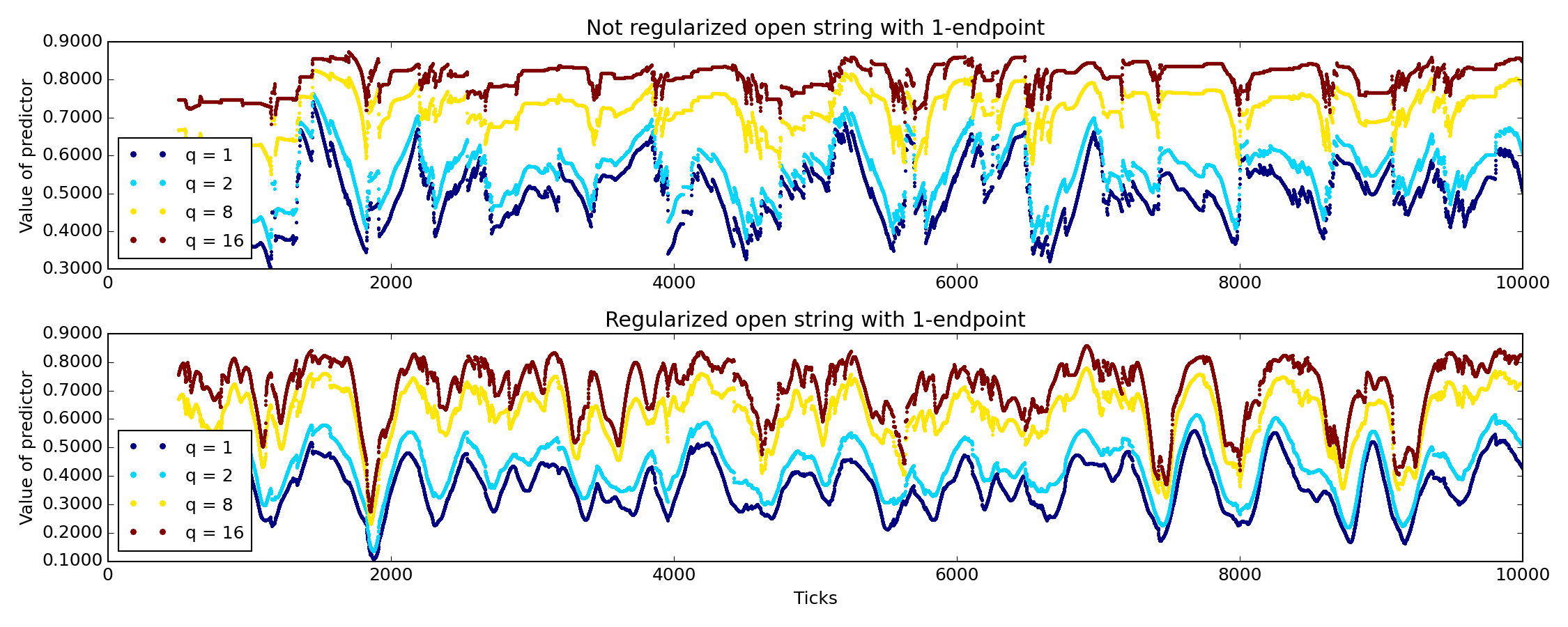}
		\label{fig:pOS1ep}
	}\\
	\subfloat[2-endpoints open string]{
		\includegraphics[width=\columnwidth]{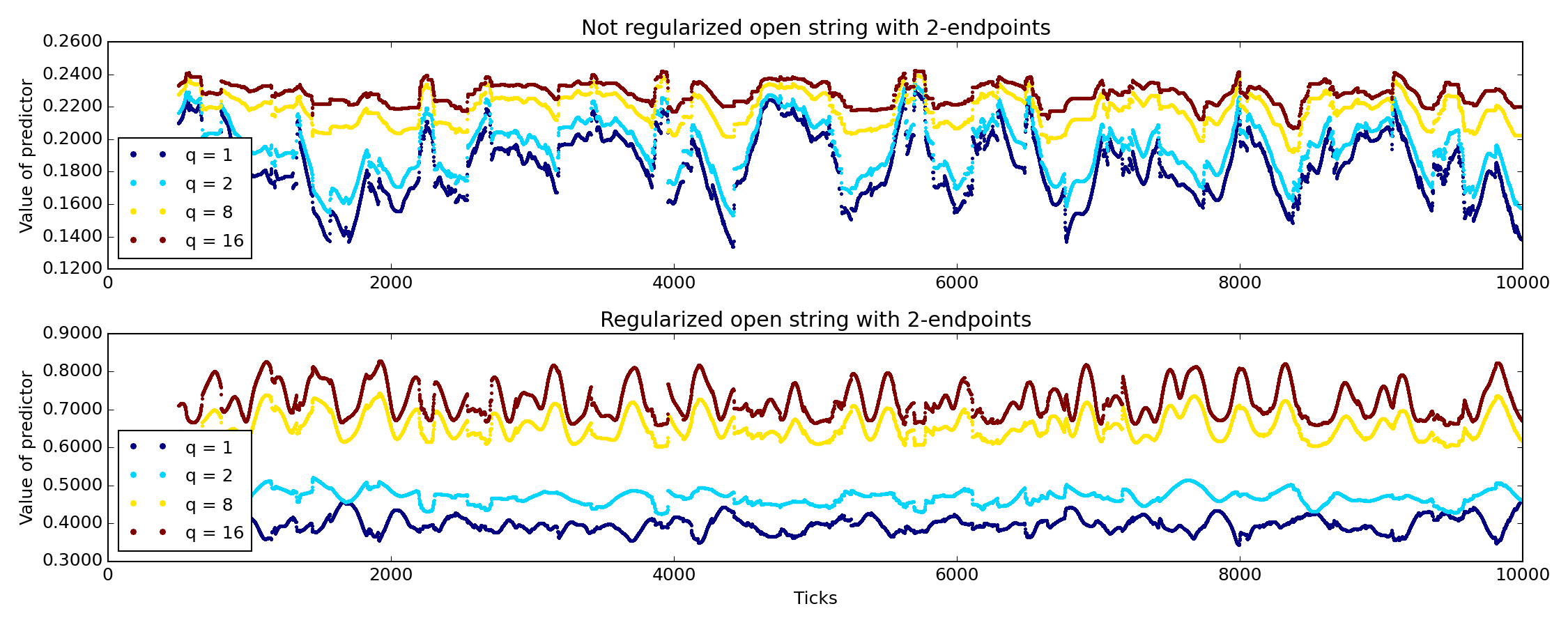}
		\label{fig:pOS2ep}
	}
	\caption{Not regularized and regularized values of the momentum of the string for the sample of time series ticks in the case of 1-endpoint open string (a) and 2-endpoints open string (b) for typical values of $q$ parameter. \label{fig:pOS}}
\end{figure}

Such model yields to the momenta values depicted in Fig.~\ref{fig:pOS1ep}. Despite the simple approach, the results of the simulations with the prediction model in the OANDA market have demonstrated the stability of the proposed trading algorithm on the transaction costs for the long trade periods. However, the values of $M$ have not allowed to find the wide regions of invariants even for the higher $q$ or other periodic functions as in Eq.~(\ref{eq:rf}), which lead us to find more complex solutions (more details in Appendix~\ref{app:B}).

\subsection{Open string with two endpoints}

In the next we study the influence of more complex string objects on the momentum behavior. At first we propose to incorporate a long-term trend by the nonlinear map corresponding to an open string with 2-endpoints
\begin{multline}  \label{eq:POS2ep}
P_{q}^{(2)}(\tau,h) = f_{q}\bigg(\bigg(\frac{p(\tau+h) - p(\tau)}{p(\tau+h)}\bigg)\\  \times\bigg(\frac{p(\tau+l_s) - p(\tau+h)}{p(\tau+l_s)}\bigg)\bigg),
\end{multline}
with $h\in \bra 0,l_s\ket$, $q$ deformation $f_{q}=\m{sign}(x)|x|^{q}$ and $p(\tau)$ value represents the mean value $p(\tau) = (p_{\m{ask}}(\tau) + p_{\m{bid}}(\tau))/2$. The $P_{q}^{(2)}(\tau,h)$ fulfills boundary conditions of Dirichlet type
\begin{equation}
P_{q}^{(2)}(\tau,0)= P_{q}^{(2)}(\tau,l_s)=0\,,\qquad
\mbox{at all ticks}\,\,\, \tau\,.
\label{eq:Dir}
\end{equation}
Practically, one can replace $P_{N}(\tau,h)\to P_{q}^{(2)}(\tau,h)$ in Eq.~(\ref{eq:mp}) and look at the values of $M$. Fig.~\ref{fig:pOS2ep} shows that the effect of the regularization is notable in comparison with previous case of 1-endpoint open string, even for low values of $q$ parameter. It allow us to focus on the predictor values which determine the stability of the algorithm or in other words they reflect the price changes on the scale of string length.

\subsection{Open polarized string with two endpoints}

Further modification of the string map to include spread-adjusted currency return $(p_{\m{bid}}(\tau) - p_{\m{ask}}(\tau))/(p(\tau,h))$ is rather straightforward, it is an analogy with a charged string polarized by an external field. The formula has the form
\begin{align} \label{eq:POPS2ep}
P_{q}^{\m{ab}}(\tau,h) =& f_{q}\bigg(\bigg(\frac{p_{\m{bid}}(\tau+h) - p_{\m{ask}}(\tau)}{p(\tau+h)}\bigg)\nn \\
&\times \bigg(\frac{p_{\m{bid}}(\tau+l_s) - p_{\m{ask}}(\tau+h)}{p(\tau+l_s)}\bigg)\bigg)\,.
\end{align}
The violation of the Dirichlet boundary condition is restored, for instance, by the subtraction $\tilde{P}_{q}^{\m{ab}}(\tau,h) \equiv P_{q}^{\m{ab}}(\tau,h) - P_{q}^{\m{ab}}(\tau,0)$. For the polarized string mapping, i.~e., the replacement $P_{N}(\tau,h)\to P_{q}^{\m{ab}}(\tau,h)$, the regularized and nonregularized values of the momenta $M$ looks identically to the previous case of open string (see Fig.~\ref{fig:pOS2ep}) and the simulations yield to the similar results.

\begin{figure}[!tb]
	\begin{center}
		\subfloat[$q = 1$]{
			\includegraphics[width=0.38\columnwidth]{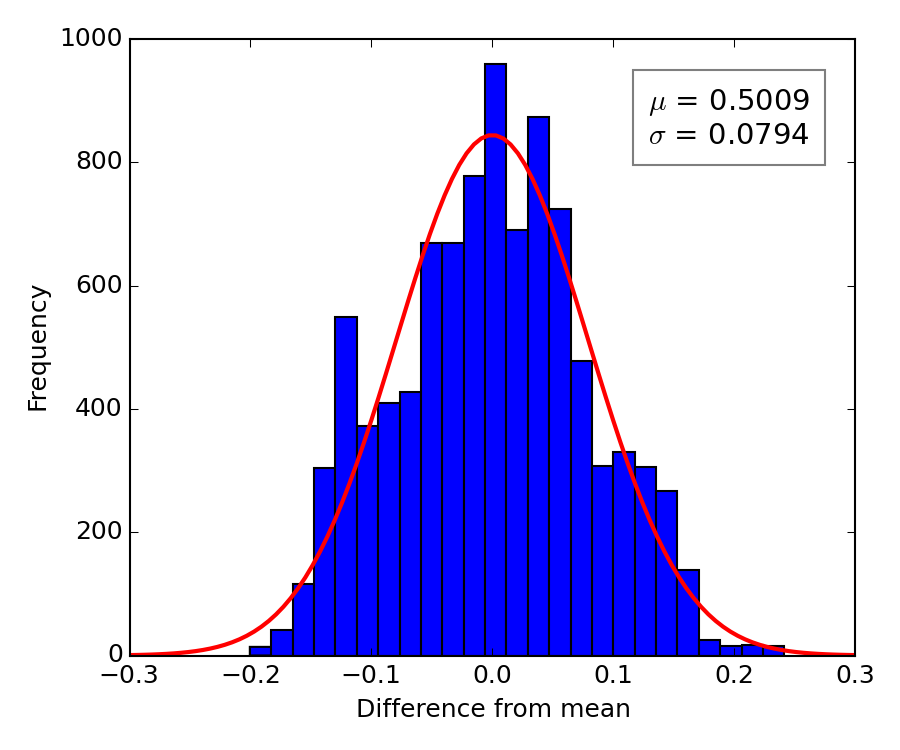}
		}
		\subfloat[$q = 1$]{
			\includegraphics[width=0.38\columnwidth]{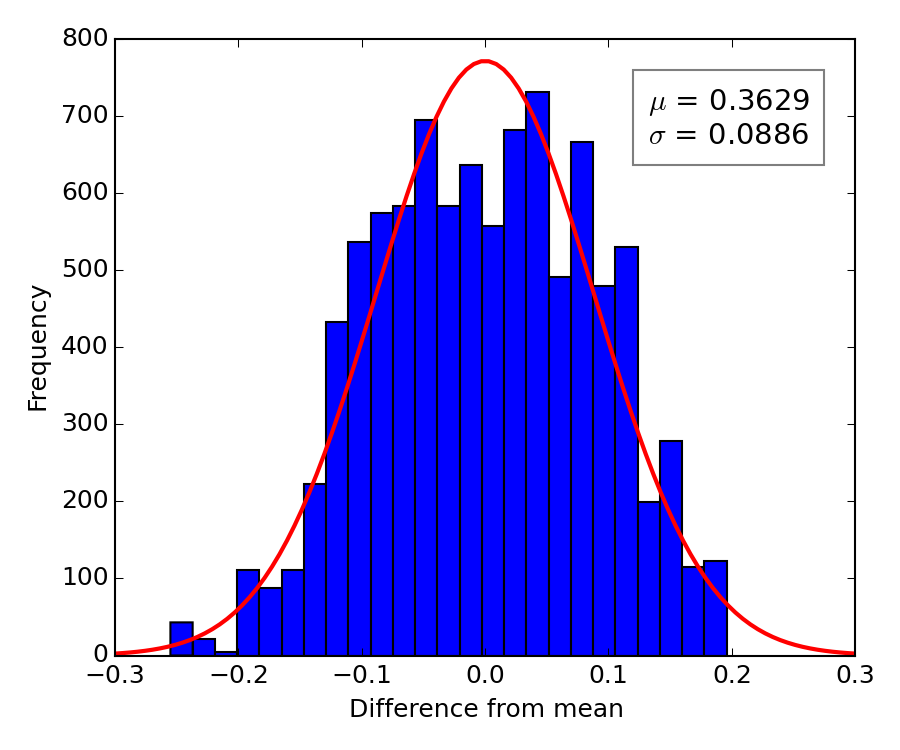}
		}\\
		\subfloat[$q = 2$]{
			\includegraphics[width=0.38\columnwidth]{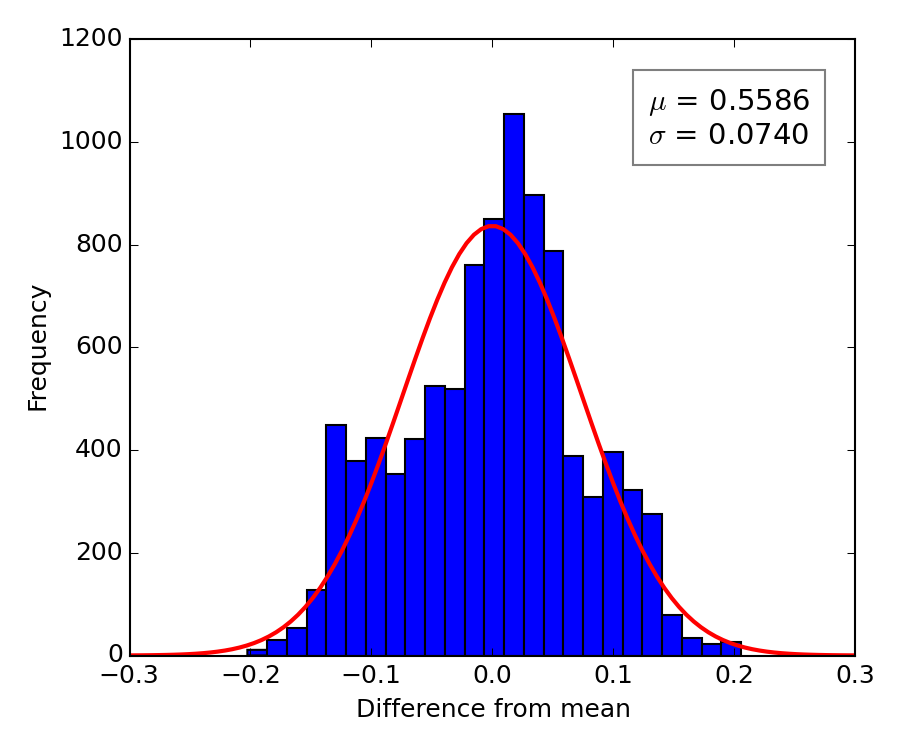}
		}
		\subfloat[$q = 2$]{
			\includegraphics[width=0.38\columnwidth]{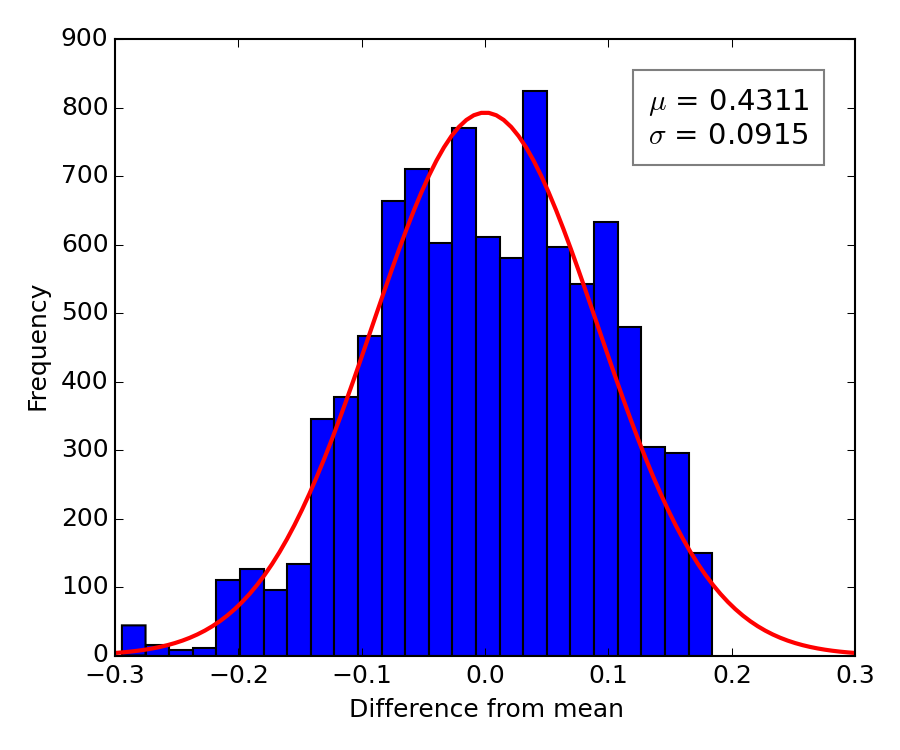}
		}\\
		\subfloat[$q = 8$]{
			\includegraphics[width=0.38\columnwidth]{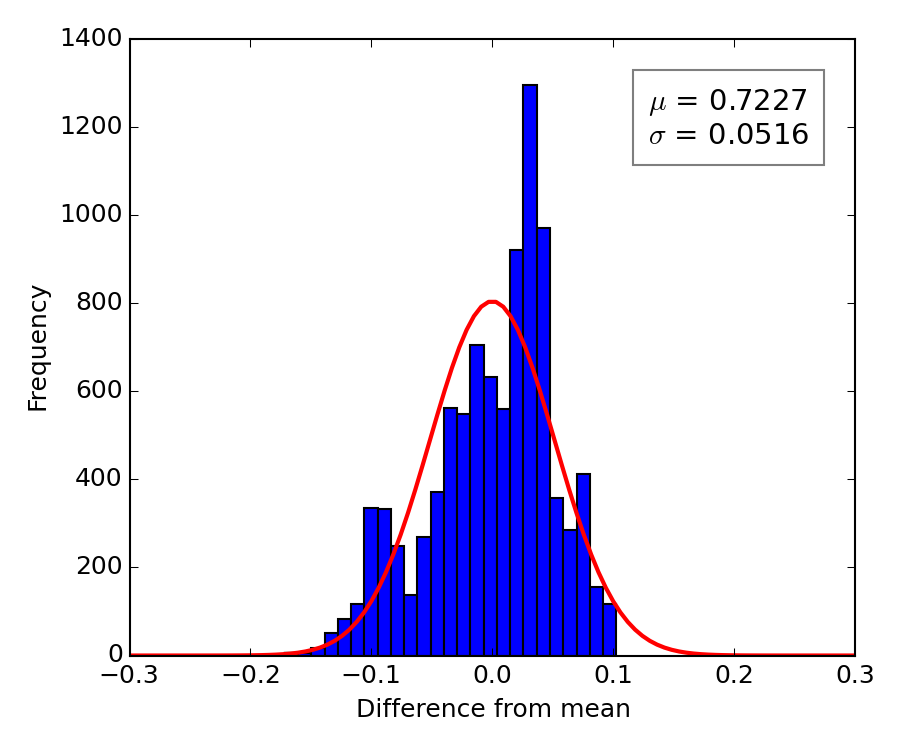}
		}
		\subfloat[$q = 8$]{
			\includegraphics[width=0.38\columnwidth]{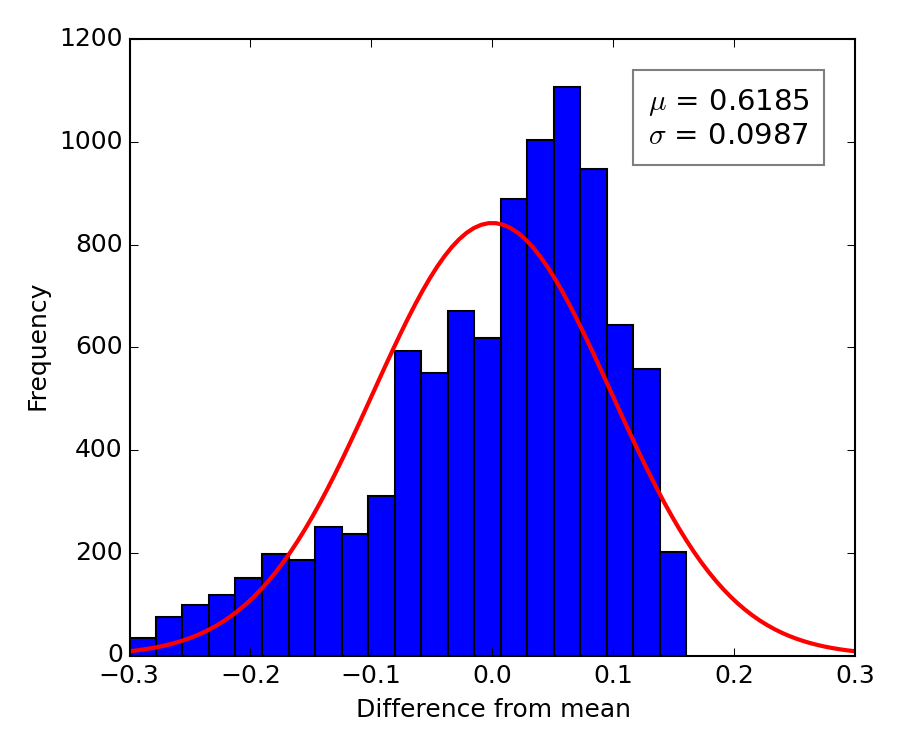}
		}\\
		\subfloat[$q = 16$]{
			\includegraphics[width=0.38\columnwidth]{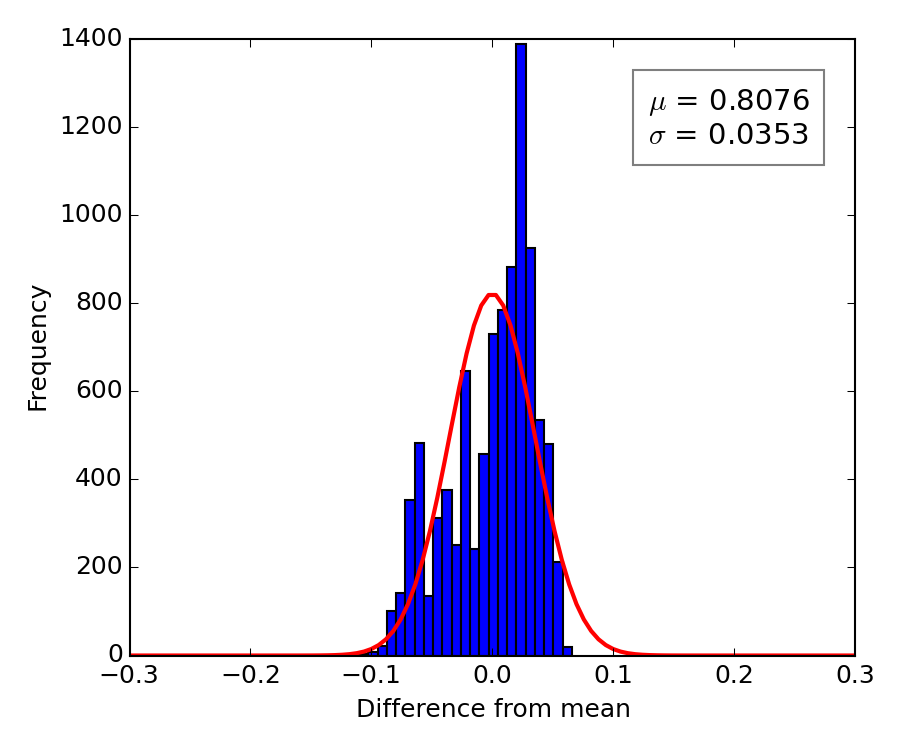}
		}
		\subfloat[$q = 16$]{
			\includegraphics[width=0.38\columnwidth]{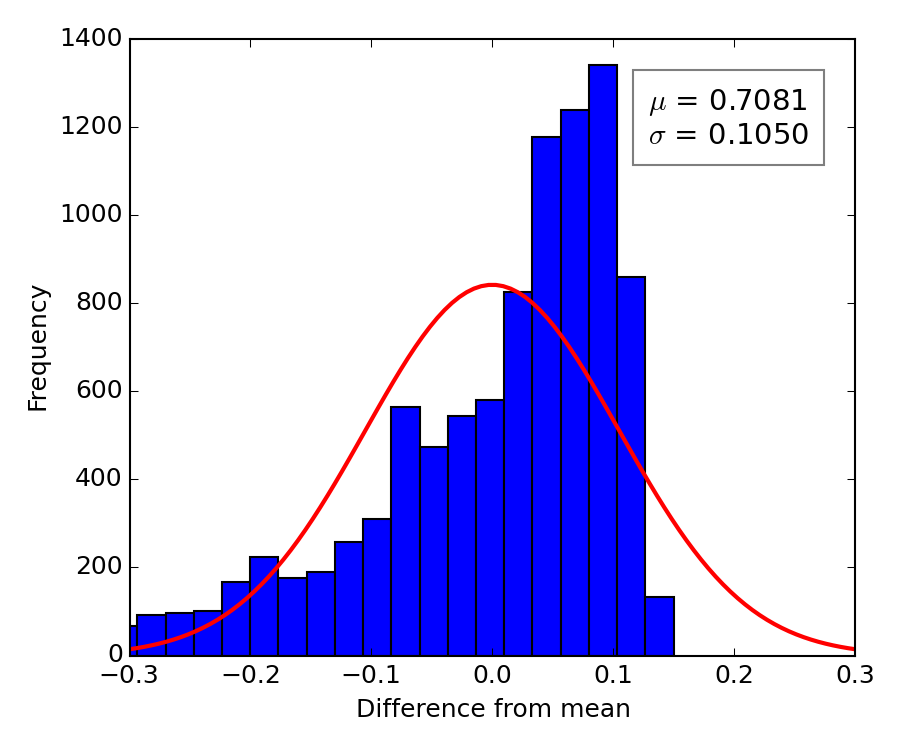}
		}
	\end{center}
	\caption{Values of momenta for 1-endpoint open string without (left column) and with (right column) regularization function. Histograms show the difference of values from the mean of normal distribution function ($\mu$, $\sigma$). \label{fig:histOS1ep}}
\end{figure}

To quantify the received predictor statistics one can construct the histograms for a spectrum of $M$ momenta as shown in Fig.~\ref{fig:histOS1ep}. Broader peaks of the distributions for regularized values of $M$ for 1-endpoint and 2-endpoints strings suggests that the values are more smoothed than in the unregularized case and the aims to forecast the market trends are based on the sharper values of $M$, i.~e., only the highest changes of a price on the market are taken into account and in this way they facilitate the evaluation of buy/sell orders.

\subsection{D2-brane model}

More interesting way how to go beyond a string model is to extent the string lines towards the more complex maps, the membranes called D2-branes. Practically it can be realized with the mapping in the form
\begin{align} \label{eq:PD2q}
P_{\m{D2},q}(\tau, h_1, h_2)  =& 
f_q\Bigg(\bigg(\frac{p^{\m{ask}}(\tau + h_1) -  p^{\m{ask}}(\tau)}{p^{\m{ask}}(\tau + h_1)}  \bigg)\nn\\
&\times\bigg(\frac{ p^{\m{ask}}(\tau+l_s) - p^{\m{ask}}(\tau+h_1)}{p^{\m{ask}}(\tau +l_s)} \bigg) \nn\\
&\times
\bigg(\frac{p^{\m{bid}}(\tau) - p^{\m{bid}}(\tau+h_2) }{p^{\m{bid}}(\tau)}\bigg) \\
&\times
\bigg(\frac{p^{\m{bid}}(\tau+h_2) - p^{\m{bid}}(\tau+l_s)}{p^{\m{bid}}(\tau+h_2)} \bigg)\Bigg) \,. \nn
\end{align}
with the coordinates $(h_1, h_2) \in \langle 0,l_{\rm s}\rangle \times \langle 0,l_{\rm s}\rangle$ which vary along two extra dimensions. The mapping satisfies the Dirichlet boundary conditions
\begin{align}
P_{\m{D2},q}(\tau, h_1, 0)   &=   P_{\m{D2},q}(\tau, h_1, l_s)   =\nn \\
P_{\m{D2},q}(\tau, 0,  h_2)  &=   P_{\m{D2},q}(\tau, l_s, h_2).
\end{align}
The momentum of D2-brane model can be modified to
\begin{multline} \label{eq:mpD2}
M(l_s, m, q, \varphi, \varepsilon) = \Bigg(\frac{1}{(l_s+1)^2}\\ \times
\sum_{h_1=0}^{l_s}\sum_{h_2=0}^{l_s}\Big| P_{\m{D2},q}(\tau,h_1,h_2) - F_{\m{D2}}(h,\varphi,\varepsilon)\Big|^q\Bigg)^{1/q},
\end{multline}
the regular function depends also on more variables, e.~g., it can has the form
\begin{align}
F_{\m{D2}}(h,\varphi,\varepsilon) = \frac{1}{2}\big(\sin(\tilde{\varphi}^2)\cos(\tilde{\varepsilon}^2)\big),\, \tilde{\varepsilon} = \frac{2\pi m h}{l_s + 1} + \varepsilon.
\end{align}

\begin{figure}[!tb]
	\begin{center}
		\includegraphics[width=\columnwidth]{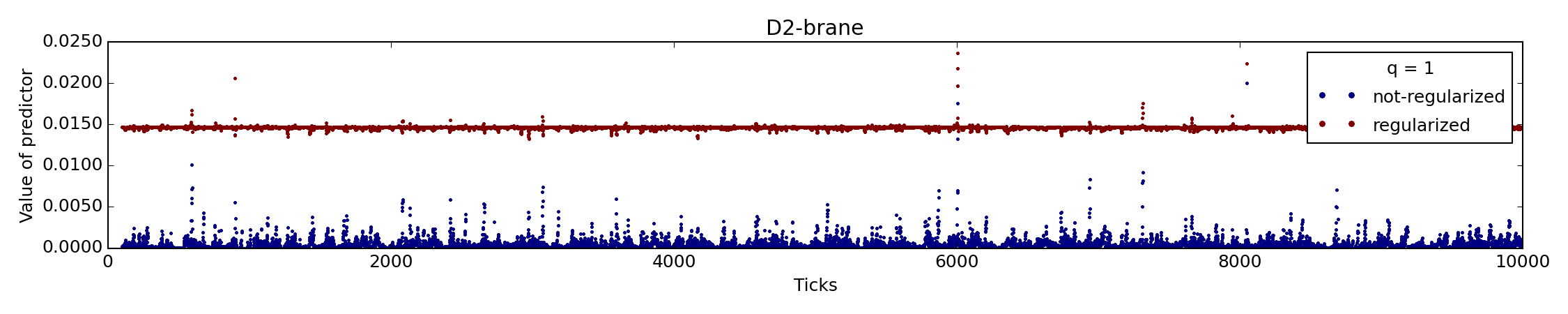}\\
		\includegraphics[width=\columnwidth]{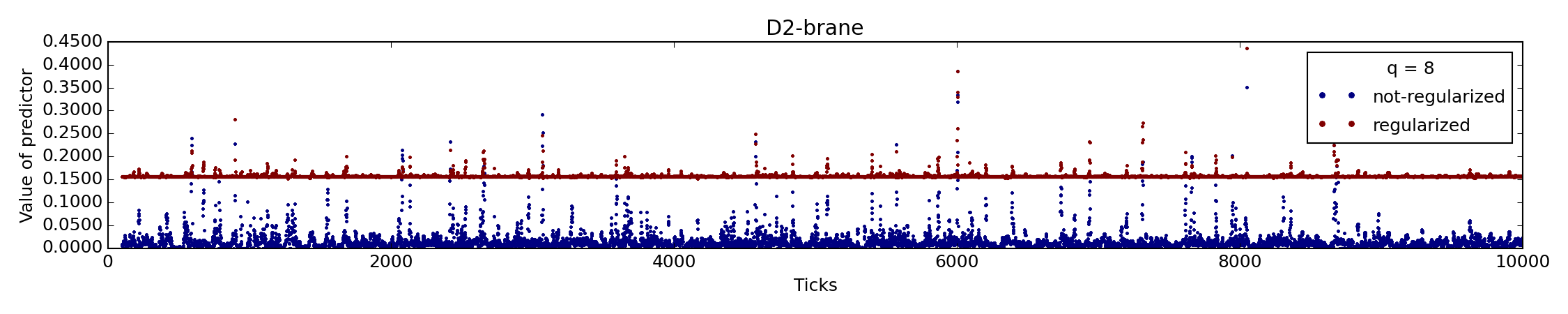}%
	\end{center}
	\caption{Not regularized (blue) and regularized (red) values of the momenta for D2-brane. The sample of 10 thousand time series ticks, $q=1$ and $q=8$. \label{fig:pD2}}
\end{figure}

\begin{figure}[!tb]
	\begin{center}
		\subfloat[$q = 1$]{
			\includegraphics[width=0.38\columnwidth]{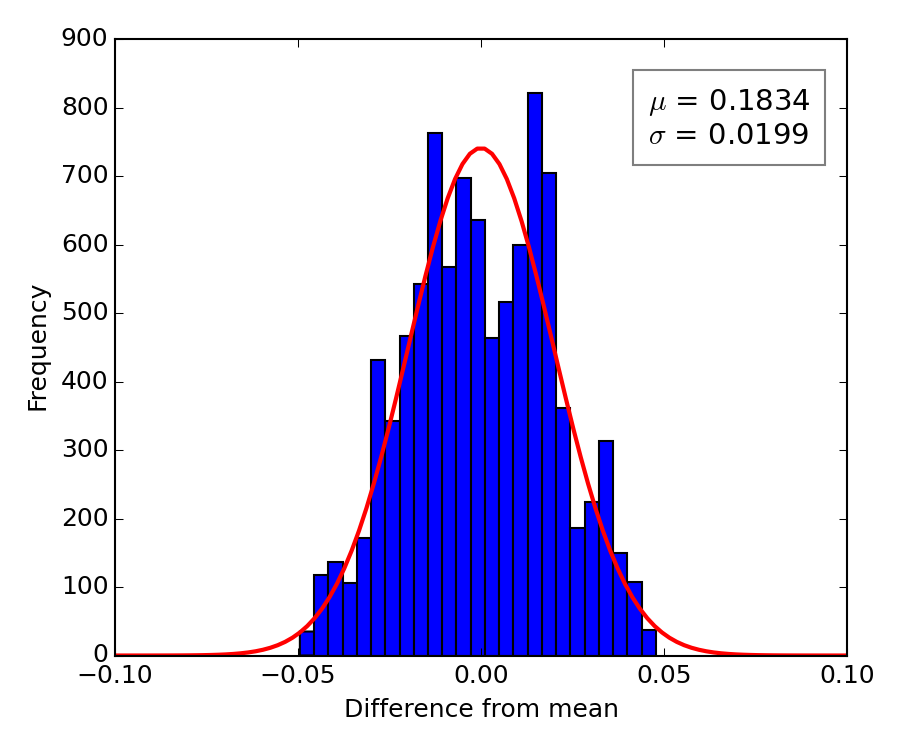}
		}
		\subfloat[$q = 1$]{
			\includegraphics[width=0.38\columnwidth]{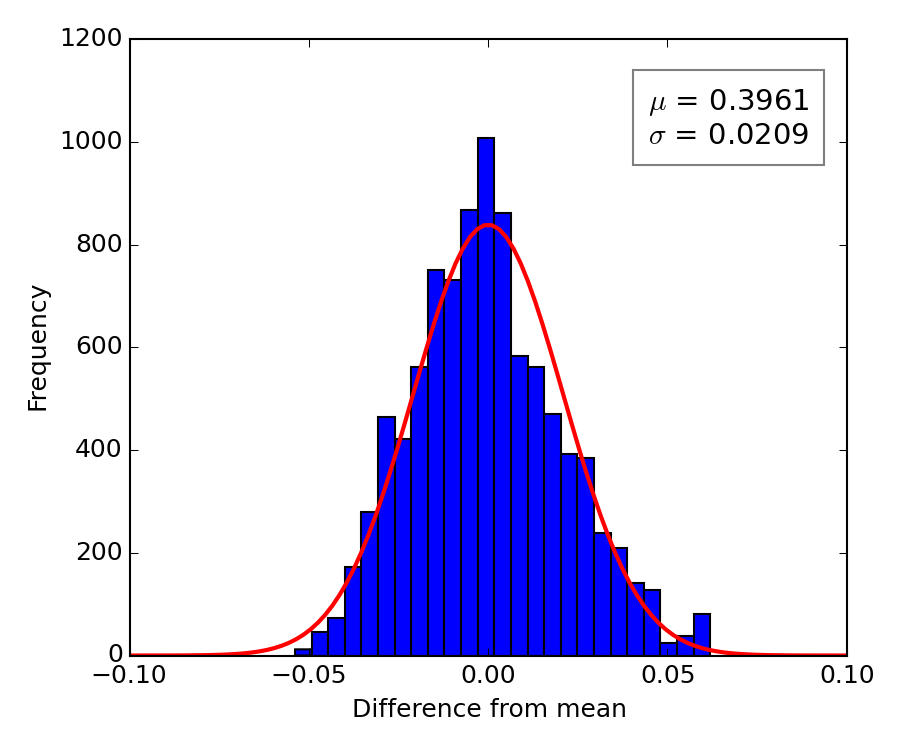}
		}\\
		\subfloat[$q = 8$]{
			\includegraphics[width=0.38\columnwidth]{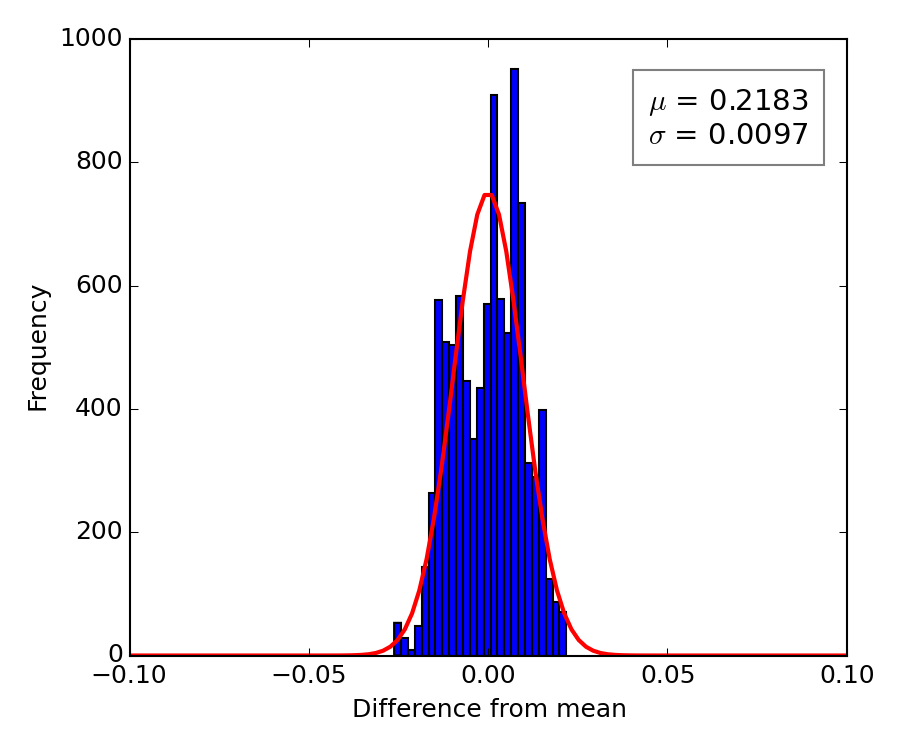}
		}
		\subfloat[$q = 8$]{
			\includegraphics[width=0.38\columnwidth]{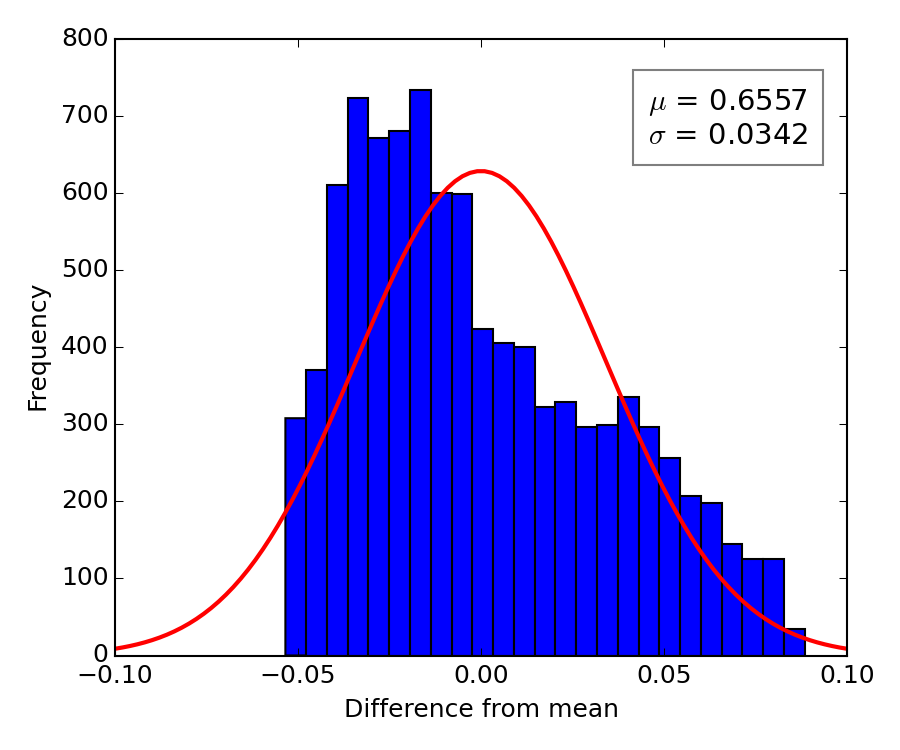}
		}\\
		\subfloat[$q = 1$]{
			\includegraphics[width=0.38\columnwidth]{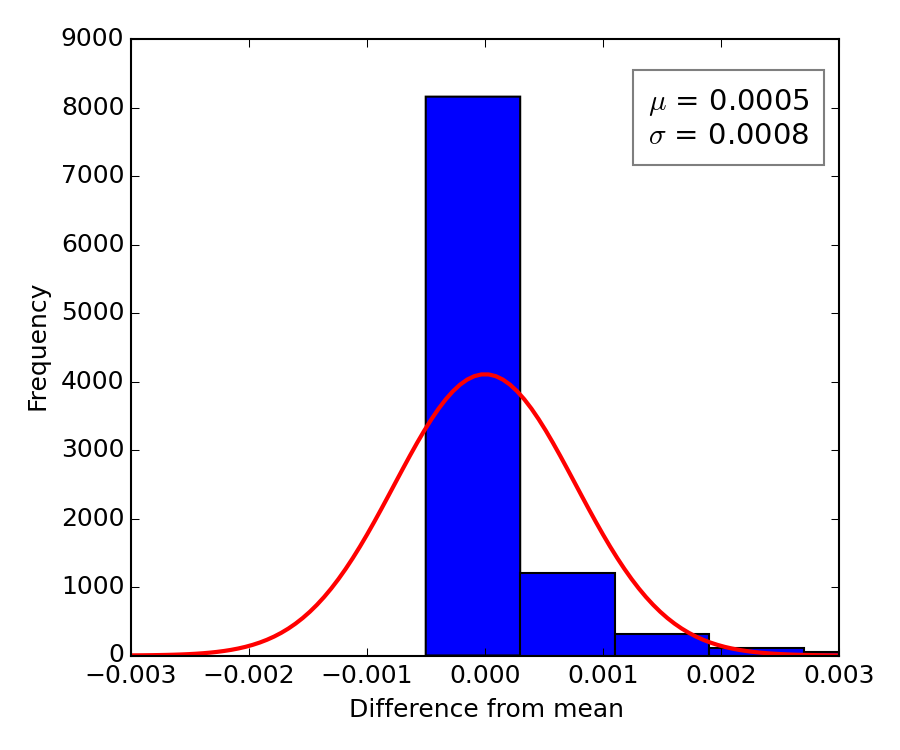}
		}
		\subfloat[$q = 1$]{
			\includegraphics[width=0.38\columnwidth]{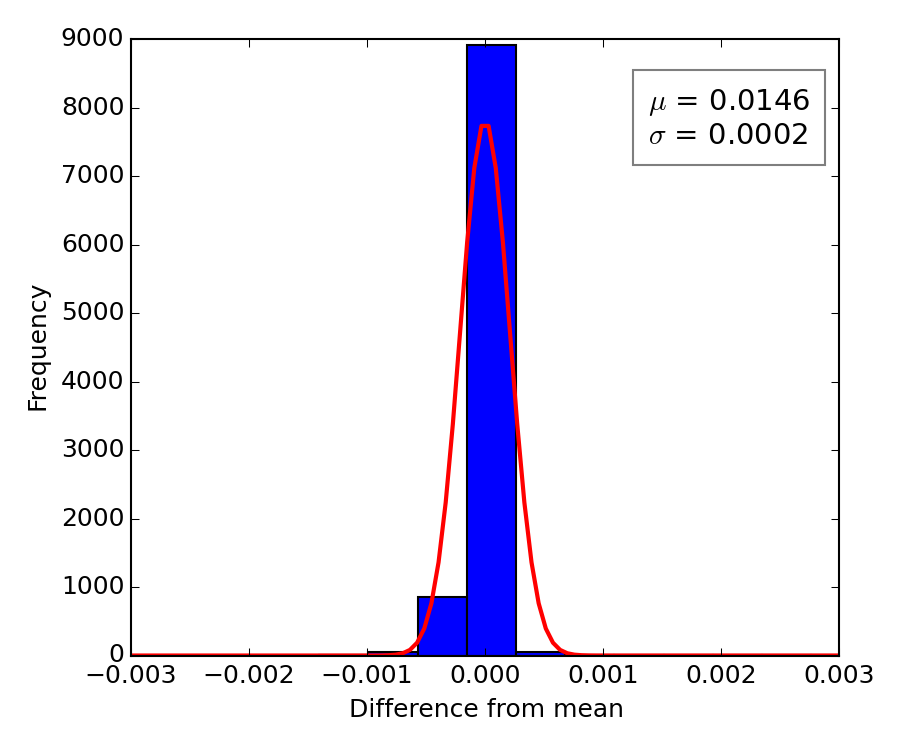}
		}\\
		\subfloat[$q = 8$]{
			\includegraphics[width=0.38\columnwidth]{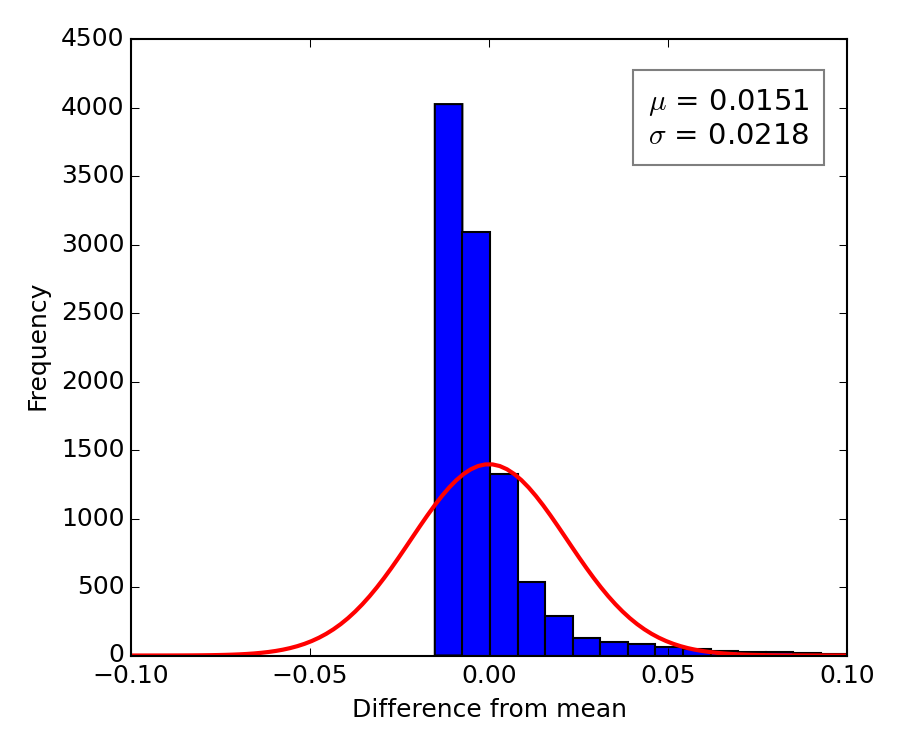}
		}
		\subfloat[$q = 8$]{
			\includegraphics[width=0.38\columnwidth]{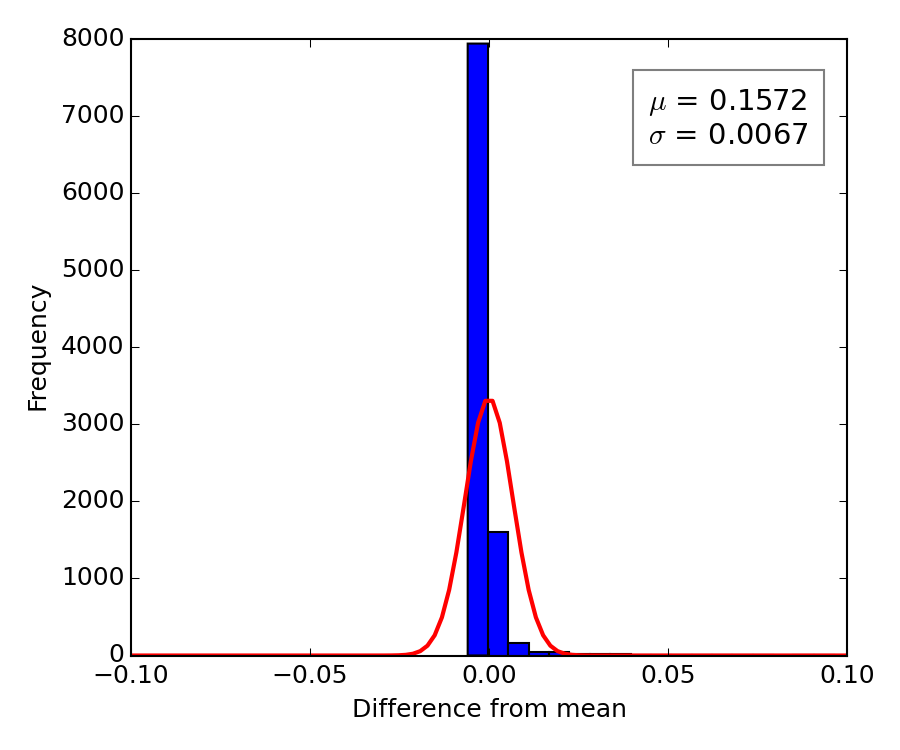}
		}
	\end{center}
	\caption{Values of momenta for 2-endpoints open string ((a)--(d)) and D2-branes ((e)--(h)) without (left column) and with (right column) regularization function. Histograms show the difference of values from the mean of normal distribution function ($\mu$, $\sigma$). \label{fig:histMix}}
\end{figure}

The effect of higher dimension D2-branes onto the $M$ values in Eq.~(\ref{eq:mpD2}) is visible in Fig.~\ref{fig:pD2}. In comparison with 1-endpoint and 2-endpoints open strings the unregularized values are more smoothed. The regularization does not improve the spectrum so significantly as in the previous case of string models as it is visible from the histograms shown in Fig.~\ref{fig:histMix}. One can conclude that even the D2-branes model with basic configuration is suitable to capture the dynamic changes of prices on the financial market.

\begin{figure*}[!tb]
	\centering
	\includegraphics[width=\columnwidth]{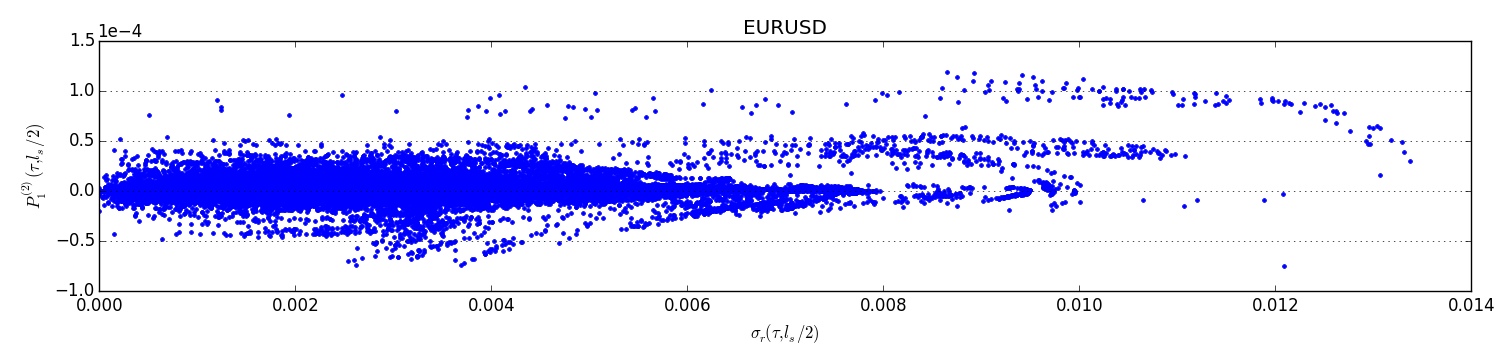}
	\includegraphics[width=\columnwidth]{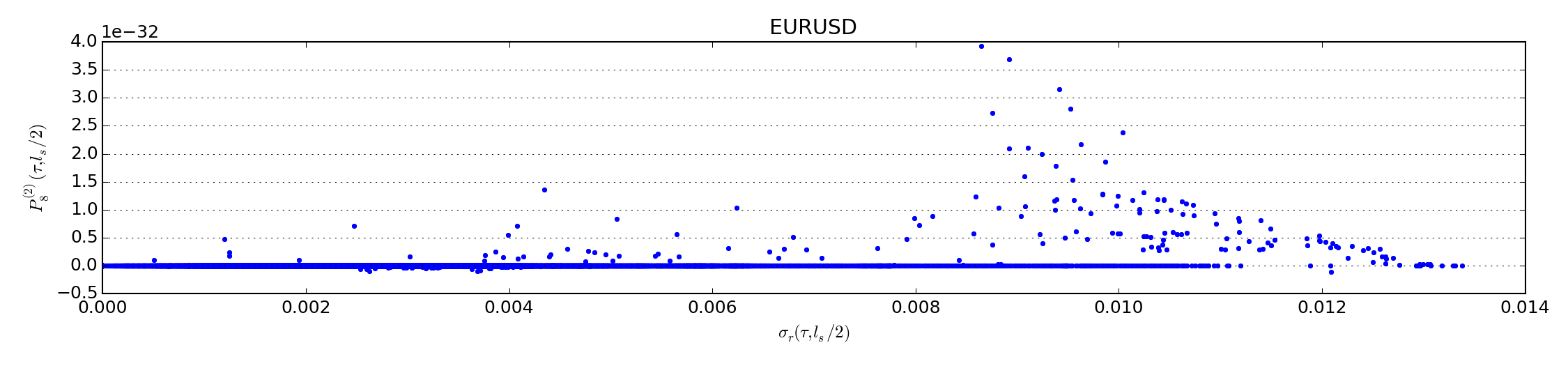}\\
	\includegraphics[width=\columnwidth]{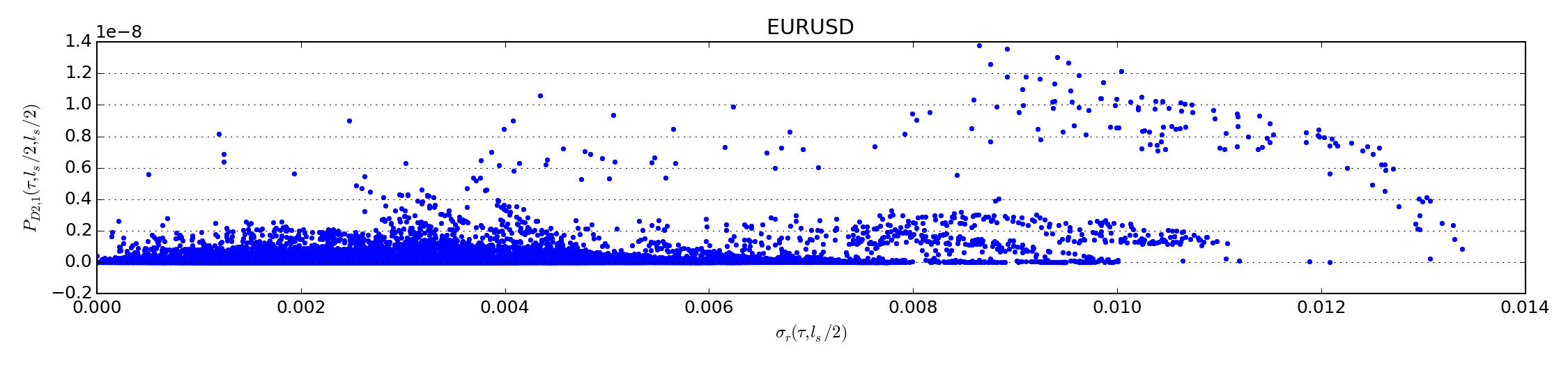}
	\includegraphics[width=\columnwidth]{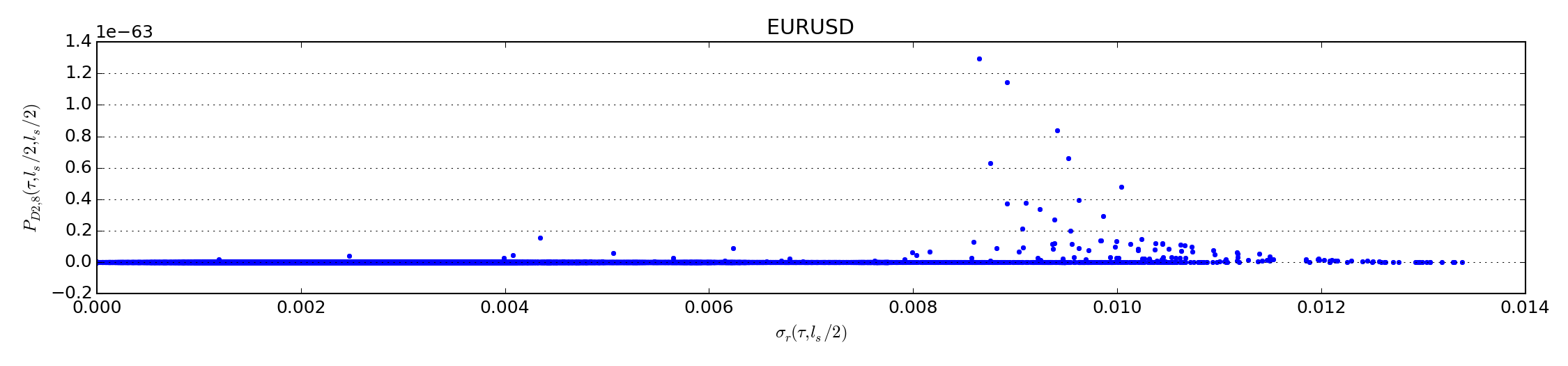}
	\caption{Relationship of return volatility $\sigma_r(l_s/2)$ and the string amplitudes for 2-endpoints string (first row) and D2-brane (second row). It shows the separating effect for $q=1$ and $q=8$. Calculated for 1min. EUR/USD ticks at time period $01-12/2015$ and $l_s=1000$. \label{fig:return}}
\end{figure*}

As another tool for evaluating of the different approaches represented by the string and D2-branes models can server the return volatility $\sigma_{ls/2}$. In contrast to a historical volatility (the standard deviation of currency returns), the return volatility acts at the time scale $l_s/2$ as string statistical characteristic. It is defined as
\begin{align}
\sigma_r(l_s/2) &= \sqrt{r_2(l_s/2) - r_1^2(l_s/2)},\\ \nn
r_m(l_s/2) &= \sum_{h=1}^{l_s/2}\big[(p(\tau+h) - p(\tau+h-1))/(p(\tau + h))\big]^{m},
\end{align}
for $m=1,2$. The scatterplot in Fig.~\ref{fig:return} shows the relationship of return volatility at the scale of $l_s/2$ to the changes in the price trends represented by the string amplitudes for 2-endpoints string $P_{i}^{(2)}(\tau,l_s/2)$ and D2-brane $P_{D2,i}(\tau,l_s/2,l_s/2)$, $i=1,8$.

The impact of high $q$ to identify the rare events of volatility is visible in both cases, nevertheless, if one decides or does not decide to use the $q$-deformed model in favor of D2-branes it depends also on the technical conditions of real time calculations, because to receive the statistics and to make predictions with D2-branes requires more computing power.

\subsection{Comparision}\label{sec:regression}

\begin{figure*}[!tb]
	\centering
	\includegraphics[width=\linewidth]{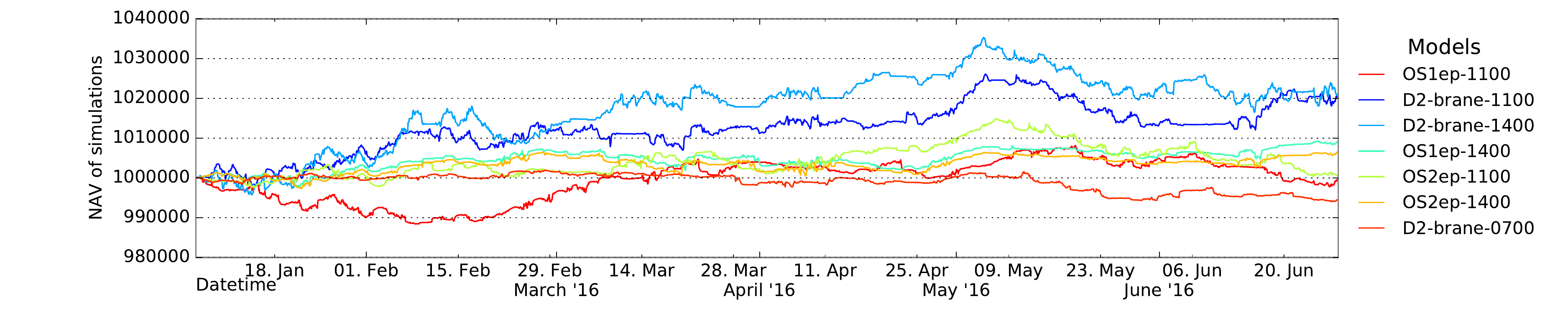}\\
	\includegraphics[width=\linewidth]{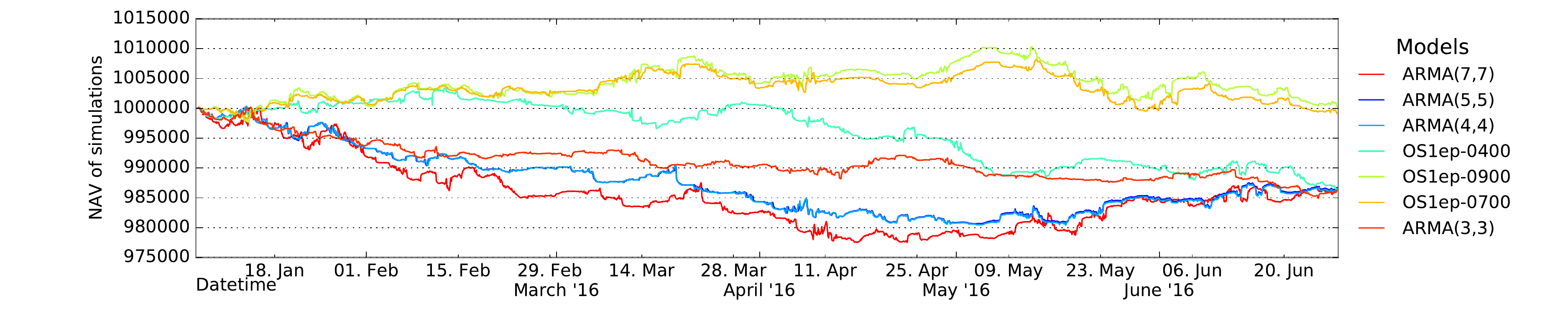}
	
	\caption{Net asset value plots for model simulations on the EUR/USD currency rate for a time period 01 -- 06/2016 as the dependence on string mapping (OS1ep -- one string with one endpoint, OS2ep -- open string with two endpoints, D2-branes model) compared to time series forecasting models of ARMA type. Number in string model denotes the value of string length $l_s$. \label{fig:comparo}}
\end{figure*}

For the purpose to demonstrate the impact of different types of string maps on the net asset value (NAV) we performed numerical simulations with open strings with one and two endpoints, D2-branes and ARMA(p,q) type forecasting models on trade online system (more in Appendix~\ref{app:B}) with build-in derived algorithms. The plot in Fig.~\ref{fig:comparo} presents the results of the simulations for EUR/USD currency pair. In the simulations we have tried to keep all parameters the same as possible, the impact of string length $l_s$ was tested on final result, OS1ep and OS2ep models have the same regularization function with $q=8$, D2-brane model is not regularized. The study revealed the incapability of ARMA models to keep even zero profit. On the contrary, the results of the string models revealed improvement of NAV with the transition from 1-endpoint to to 2-endpoints open string and D2-branes. Moreover, the higher efficiency for the string models may be achieved by longer string $l_s$ lengths.

\section{Regge slope parameter}\label{sec:regge}

In this section we closely look at another quantity which has origin in the string theory, so called Regge slope parameter $\alpha'$. The connection of the slope parameter and the angular momentum makes it suitable for the investigation of the stability of currency rates as shown below.

For rotating open string, the parameter $\alpha'$ or inverse of the string tension, is the constant that relates the angular momentum of the string $J$ to the square of its energy $E$
\begin{equation}
\alpha'=\frac{J}{\hbar E^2}.
\end{equation}
In our analogy we introduce the slope parameter in terms of the angular momentum $M_{q}^{\m{ab}}(\tau)$.

For the time series of open-high-low-close (OHLC) values of currency rates $p(\tau)$ one can construct separated ask and bid strings, in our case we use the open string with 2-endpoints and string length $l_s$, introduced via the nonlinear map in Eq.~(\ref{eq:POS2ep}). Then the momentum distance function $d_{q}^{\m{ab}}(\tau)$ between the ask string $P_{q,\m{ask}}^{(2)}(\tau,h) \equiv P_{q}^{(2)}(\tau,h)\big|_{p\to p_{\m{ask}}}$ and bid string $P_{q,\m{bid}}^{(2)}(\tau,h) \equiv P_{q}^{(2)}(\tau,h)\big|_{p\to p_{\m{bid}}}$ has the form
\begin{gather}
d_{q}^{\m{ab}}(t) = \frac{1}{l_s+1} \sum\limits_{h=0}^{l_s}
\Big| P_{q,\m{ask}}^{(2)}(\tau,h) - P_{q,\m{bid}}^{(2)}(\tau,h) \Big|.
\end{gather}
In case of rotating open string, the nonvanishing component of angular momentum is $M_{12}$, and its magnitude is denoted by $J=|M_{12}|$ (more in \cite{Zwiebach:2009})
\begin{equation}
M_{12}=\int_{0}^{\sigma_1}(X_1P_{2}^{\tau} - X_2P_{1}^{\tau}) \dd\sigma
\end{equation}
for space and conjugate components $P_{i}$, $X_{i}$, $i=1,2$ and $\sigma_1=E/T_{0}$. It leads to the relation connecting the slope parameter $\alpha'$ and $T_{0}$ as the string tension $T_{0}$
\begin{equation} \label{eq:aT}
T_{0} = \frac{1}{2\pi\,\alpha'\,\hbar c}.
\end{equation}
In our notation, the angular momentum can be written as
\begin{multline}
M_{q}^{\m{ab}}(\tau) = \sum\limits_{h=0}^{l_s} \Big[P_{q,\m{ask}}^{(2)}(\tau,h)X_{q,\m{bid}}^{(2)}(\tau,h) \\ -P_{q,\m{bid}}^{(2)}(\tau,h)X_{q,\m{ask}}^{(2)}(\tau,h)\Big]
\end{multline}
with the conjugate variable $X_{q}^{(2)}(\tau,h)$ received by the recurrent summation from $P_{q}^{(2)}(\tau,h)$ in Eq.~(\ref{eq:POS2ep}), following the relation $\dot X_{q}^{(2)}(\tau,h) = P_{q}^{(2)}(\tau,h)$
\begin{multline}
X_{q}^{(2)}(\tau,h+1) = X_{q}^{(2)}(\tau,h) \\
+ P_{q}^{(2)}(\tau,h-1)[t(\tau+h)-t(\tau + h -1)],
\end{multline}
$t(\cdot)$ denotes the timestamp for the time $\tau$ (see \cite{Horvath:2012zz}). The slope parameter has final form
\begin{align}
\alpha_{q}' &= \frac{\bra\ack M_{q}^{\m{ab}}(\tau)\ack\ket}{2\pi l_s^2}.
\end{align}
Table~\ref{tab:tension} presents the typical values of slope parameter $\alpha_{q}'$ together with the mean of string amplitude $P_{1}^{(2)}(l_s/2)$ for 2-endpoints string mapping, the string tension $T_0$ is estimated with the help of Eq.~(\ref{eq:aT}) ($\hbar c = 1$). It is obvious that each currency pair operates with the own characteristic inter-string values. From the theory of D-branes is known generalized formula for the tension of $D_{p}$-brane \cite{Johnson2003d}
\begin{equation}
T_{D_{p}} = \frac{1}{g_s(2\pi)^p\, l_s^{p+1}},
\end{equation}
$l_s$ is the familiar string length and $g_s$ is the string coupling, which can be used for higher dimensions not considered in this work.

\begin{table*}[thbp]
	\centering
	\setlength{\tabcolsep}{10pt}
	\begin{tabular}{c c c c}
		\toprule
		\textbf{Currency pair } & $\bm{\bra P_{1}^{(2)}(l_s/2)\ket}$ & $\bm{\alpha_{1}'}$ & \ $\bm{T_0}$\ \\
		 & $[\times10^{-7}]$ & $[\times10^{-13} (2\pi)^{-1}]$ &  $[\times10^{12}]$\\
		\midrule
		AUD/CAD & \phantom{$-2$}$3.6841$  & \phantom{$1$}$8.9764$ & $1.1140$ \\
		EUR/USD & \phantom{$-2$}$0.3539$  & \phantom{$1$}$2.1890$ & $4.5684$ \\
		GBP/USD & \phantom{$2$}$-5.0099$  & \phantom{$1$}$5.4474$ & $1.8357$ \\
		USD/CAD & \phantom{$-2$}$8.6794$  &             $12.0247$ & $0.8316$\\
		USD/CHF & \phantom{$-$}$10.6082$  &             $10.6185$ & $0.9418$ \\
		USD/JPY & \phantom{$-$}$28.2180$  & \phantom{$1$}$6.9397$ & $1.4410$ \\
		\bottomrule
	\end{tabular}
	\caption{Average values of string amplitude of 2-endpoints string $P_{1}^{(2)}(l_s/2)$, slope parameter $\alpha_{1}'$ and tension $T_0$ for main currency pairs. One month (02/2016) tick data with \unit[1]{min.} resolution, string length $l_s=1000$.}\label{tab:tension}
\end{table*}

\begin{table}[thbp]
	\centering
	\setlength{\tabcolsep}{10pt}
	\begin{tabular}{c c c c c}
		\toprule
		\textbf{\ String\ } & \multicolumn{4}{c}{\textbf{\ Volatility window [in min.]}\ } \\
		\textbf{\ length\ } & \textbf{\phantom{0}5} & \textbf{10} & \textbf{40} & \textbf{60}\\
		\midrule
		10 & 0.5175 & 0.6072 & 0.3759 & 0.3531 \\
		20 & 0.3949 & 0.4574 & 0.4307 & 0.3865 \\
		30 & 0.4726 & 0.5022 & 0.5398 & 0.4935 \\
		40 & 0.4460 & 0.5098 & 0.6579 & 0.5843 \\
		50 & 0.4384 & 0.4184 & 0.5230 & 0.5240 \\
		\bottomrule
	\end{tabular}
	\caption{The Pearson product-moment correlation coefficients between the angular momentum (dependent on a string length $l_s$) and the historical volatility (dependent on a time window) calculated for close ask \unit[1]{min.} ticks of EUR/USD exchange rate on December 4th, 2015.}\label{tab:corr}
\end{table}

\section{Discussion and conclusions} \label{sec:con}

\begin{figure}[!thb]
	\centering
	\includegraphics[width=\columnwidth]{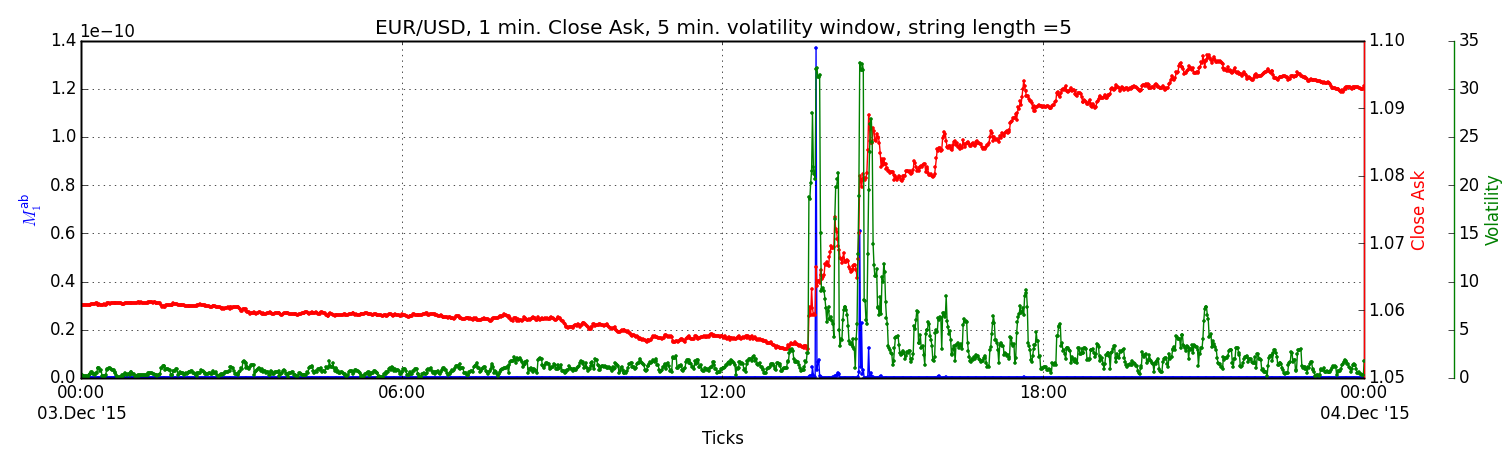}\\
	\includegraphics[width=\columnwidth]{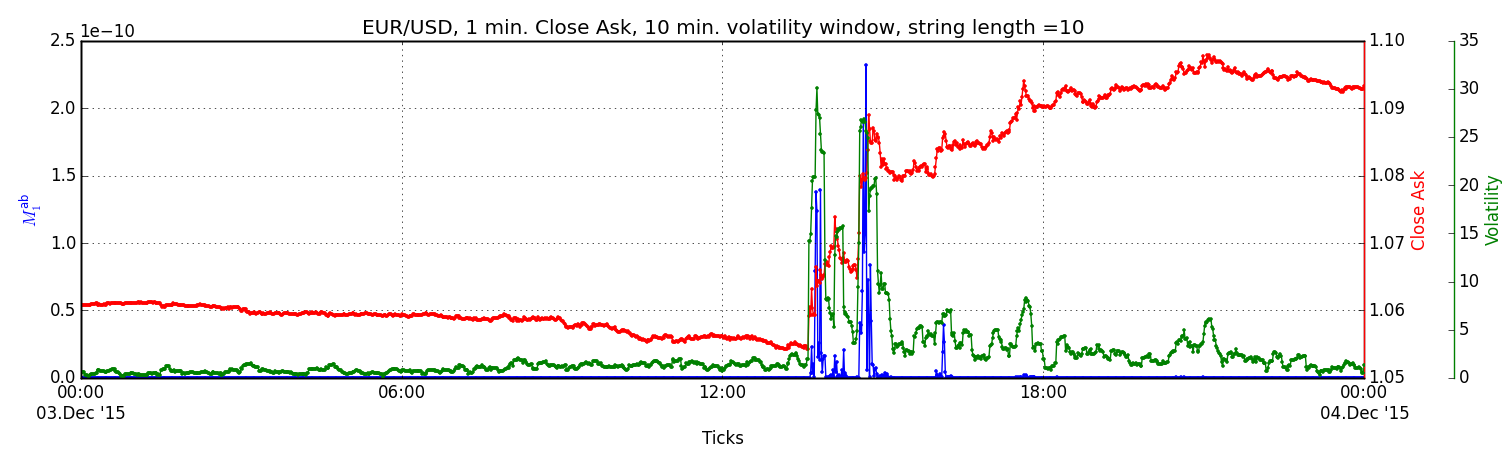}\\
	\includegraphics[width=\columnwidth]{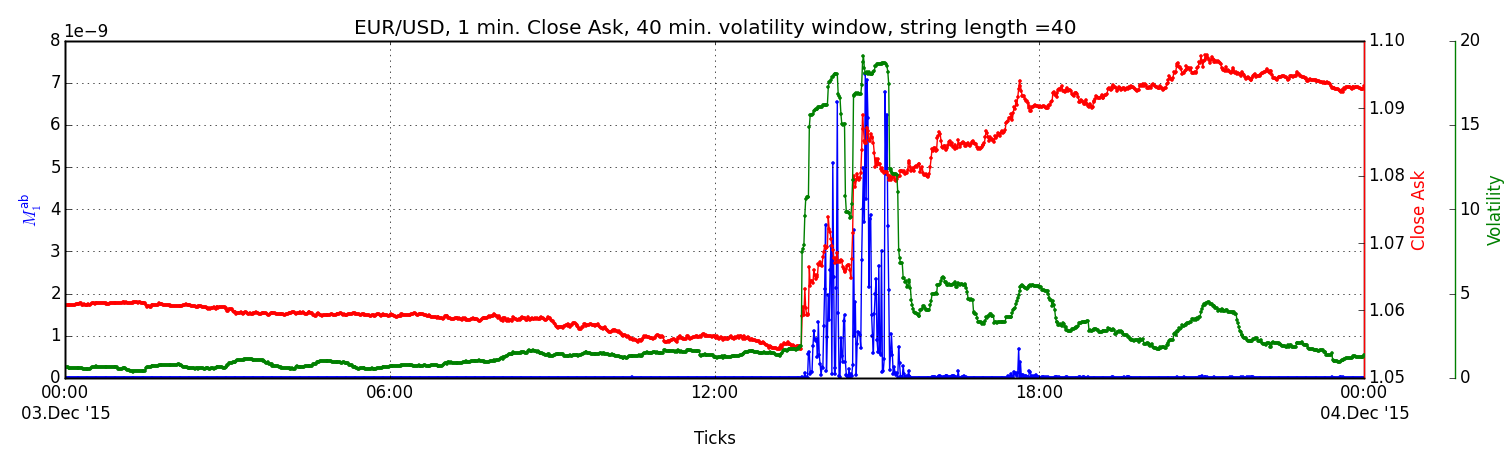}\\
	\includegraphics[width=\columnwidth]{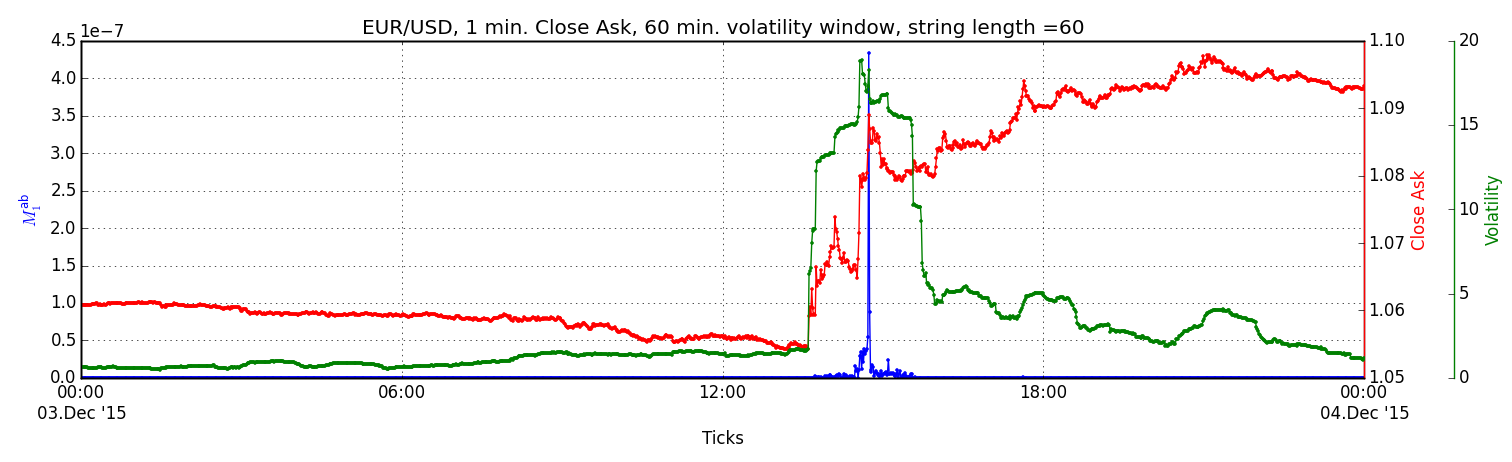}
	\caption{Plot shows the close ask value of EUR/USD exchange rate (red) for \unit[1]{min.} ticks on December 4th, 2015. The historical volatility in 5, 10, 40, 60 min. windows (green) is compared with the angular momentum $M_{q}^{\m{ab}}(\tau)$ (blue) for $q=1$ and $l_s=10$ min. \label{fig:regge1}}
\end{figure}


In the study we have introduced new string mappings to transform the currency quotes to multidimensional string objects, represented by open strings with 2-endpoints and D2-branes. The proposed objects enhance string model algorithm \cite{Pincak:2015hha} used in the real market conditions on the online trade system. We have investigated the influence of not regularized and regularized mappings on the final spectrum of momenta values for the objects (Figs.~\ref{fig:pOS}, \ref{fig:pD2}). The regularization have been obtained by the addition of the regular function to the original mapping function and $q$ parameter for the deformation. The effect of flattening of momenta values for 2-endpoints strings is notable even for low $q$ values (Fig.~\ref{fig:return}), in contrast to the 1-endpoint string, where the effect is visible only for high $q$ values. For D2-branes is the situation more favorable, the flattening is achieved already for not regularized momenta values, due to the properties of brane mapping itself. For completeness, we have mentioned also open polarized strings with 2-endpoints, but the obtained statistics is very similar to previous mappings and we have not investigated it in detail.  According the obtained corresponding numerical simulations (Fig.~\ref{fig:comparo}), the improvement of NAV with the proposed models is significant according to 1-endpoint open strings, moreover the dependence of results on string length suggests the possibility to optimize the parameters through parallel computations or evolutionary algorithms.

We also propose to apply the values of angular momentum $M_{q}^{\m{ab}}(\tau)$ as complementary tool to analyze the stability of currency rates except the historical volatility, as they are compared in Fig.~\ref{fig:regge1}. For short string lengths the angular momentum indicates the same sharp changes in exchange rate. The correlation between measures is highest for equal values of a string length and time window parameters, see Table~\ref{tab:corr}. Although there exists a certain relation between those measures, for instance, the similar sensitivity in time, the memory effect of angular momentum seems to be lower. Therefore, it may provide a helpful indication of market changes or to serve as a trade brake in algorithms.

In connection with a slope parameter $\alpha'$ and a string tension $T_0$ we have compared their values for a set of currency pairs in Table~\ref{tab:tension} (we have chosen the main six trading pairs). One can deduce that an increase of slope parameter values (or decrease of tension) indicates the changes on a market and a volatility is increasing. Although the fall in prices can last for a short time, the trading algorithms must immediately respond on the situation to avoid large losses. In Appendix~\ref{app:B} we have outlined the possible way how to deal with identified trends in real conditions. Moreover, each currency pair needs special treatment, which raises the requirement of parallel computing with genetic algorithms. We leave this as an open question for future work.

A combination of theoretical models based on geometrical description with the current financial data-driven disclosures may lead to serious revelation in traditional econometric methods. The models based on D2-branes mapping can serve as the starting point for the next generation of optimization of trade parameters in the evolutionary processes.

\appendix

\section{Sharpe ratio}\label{app:A}
	
The Sharpe ratio for calculating risk-adjusted return
\begin{equation}\label{eq:sr}
S=\frac{\mathrm{E}[r_a-r_f]}{\sigma}=\frac{\mu-r_f}{\sigma},
\end{equation}
where $r_a$ is asset return, $r_f$ is risk free rate of return, $\mathrm{E}[r_a]$ is mean asset return, $\mathrm{E}[r_a-r_f] = \mu - r_f$ is the expected value of the excess of the asset return over the benchmark return with standard notation
\begin{equation}
\Big\{x_i\Big\}_{i=1}^{N},\; \mu=\frac{1}{N}\sum\limits_{i=1}^{N}x_i,\; \sigma = \sqrt{\frac{1}{N}\sum\limits_{i=1}^{n}(x_i - \mu)^2}
\end{equation}
$\mu$ is the mean, $\sigma$ is the standard deviation.

The Sharpe ratio formula for the modified value at risk
\begin{equation}\label{eq:mvar}
S_{\mathrm{MVaR}}=\frac{\mu-r_f}{\mathrm{MVaR}},
\end{equation}
with 
\begin{align}
\mathrm{MVaR} =& -(\mu + \sigma z_{cf}), \nn\\ \nn
z_{cf} =& z_c+\frac{1}{6}\Big[(z_c^2-1)S\Big]+\frac{1}{24}\Big[(z_c^3-3z_c)K\Big]\\
&-\frac{1}{36}\Big[(2z_c^3-5z_c)S^2\Big],
\end{align}
$z_{c}$ is the $c$-quantile of the standard normal distribution, $S$ is the skewness of asset return and $K$ is the excess kurtosis of asset return.


\section{Utilization of the model for real trading}\label{app:B}

To demonstrate the behavior of physical ideas under the real trading conditions, we have constructed the trading algorithm, which has been intensively developed and tested. The algorithm version StringAlgo v.15 has demonstrated the financial forecasting on the OANDA real data for the PMBCS model \cite{Pincak:2015hha}. To see the differences and perspectives of the proposed model from the Section~\ref{sec:strings}, we present the real, i.~e., not theoretical, results from first demo sessions on the Interactive Brokers (IB) and LMAX Exchange (LMAX) market accounts, which were done through the Librade online trade system \cite{IB, LMAX, Librade}. The chosen currency pairs EUR/USD, CHF/JPY, AUD/CAD, AUD/JPY were simulated with new algorithm version and thereafter compared with the results from demo sessions (see Tab.~\ref{tab:demo1}). The algorithm StringAlgo v.16 has builtin new proposed string maps Eq.~(\ref{eq:POS2ep}), (\ref{eq:POPS2ep}), (\ref{eq:PD2q}), as well the modified Sharpe ratio (Appendix~\ref{app:A}) which serves as new statistical quantity to evaluate the value at risk. 

\begin{figure*}[hbp]
	\centering
	\includegraphics[width=1.5\columnwidth]{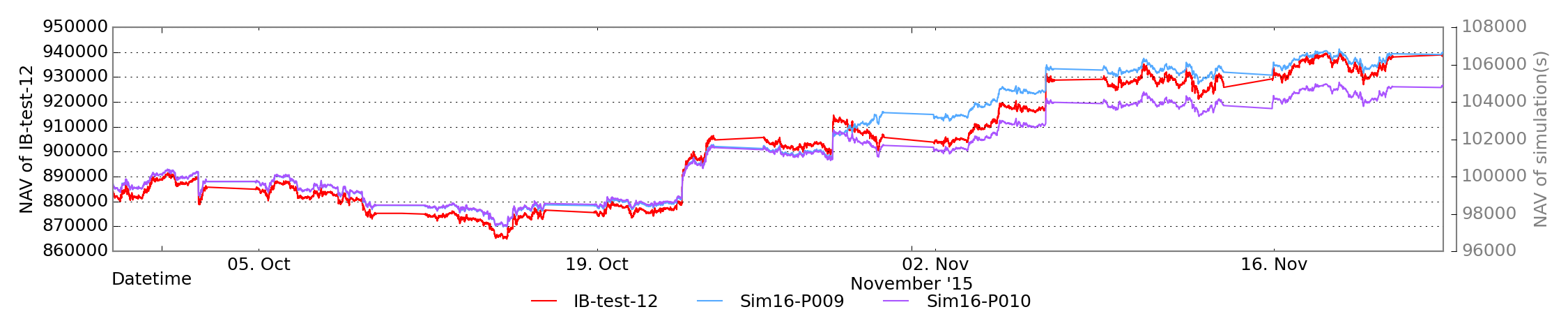}\\
	\includegraphics[width=1.5\columnwidth]{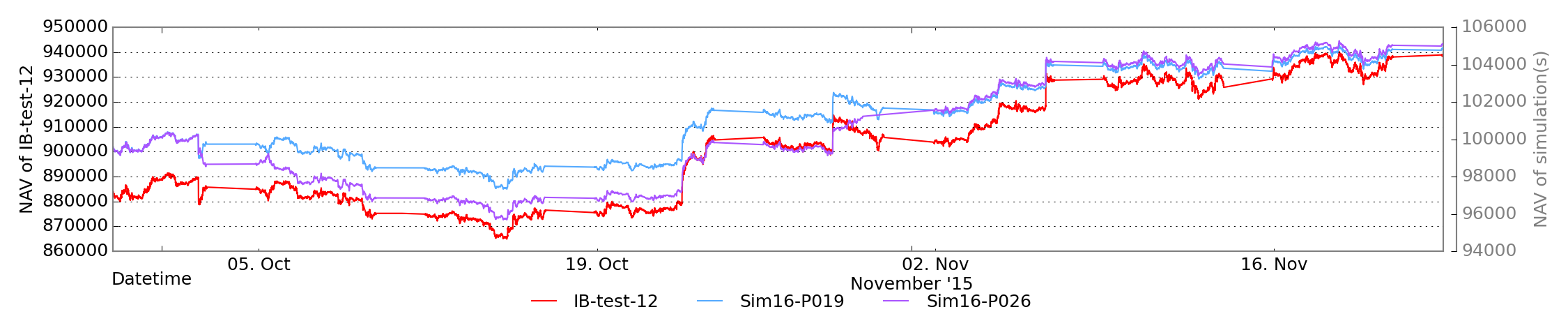}\\
	\includegraphics[width=1.5\columnwidth]{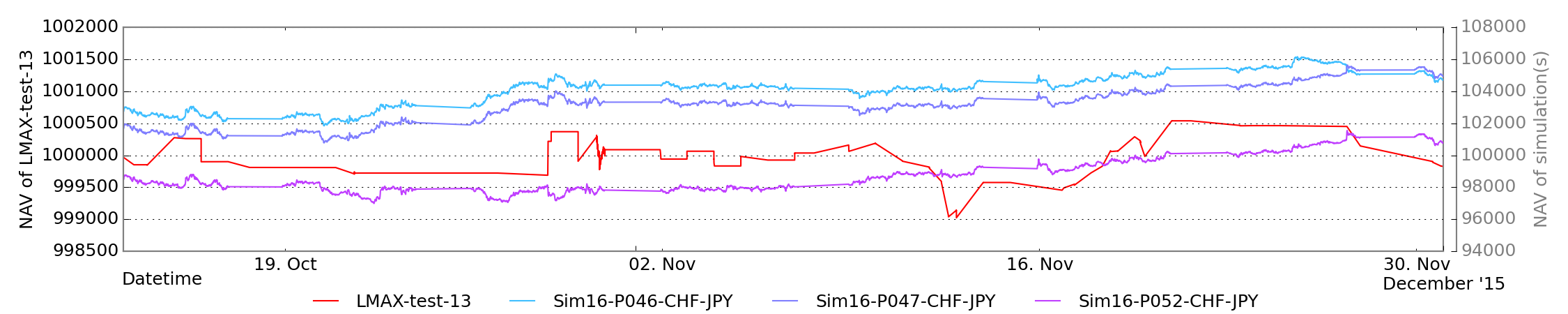}\\
	\includegraphics[width=1.5\columnwidth]{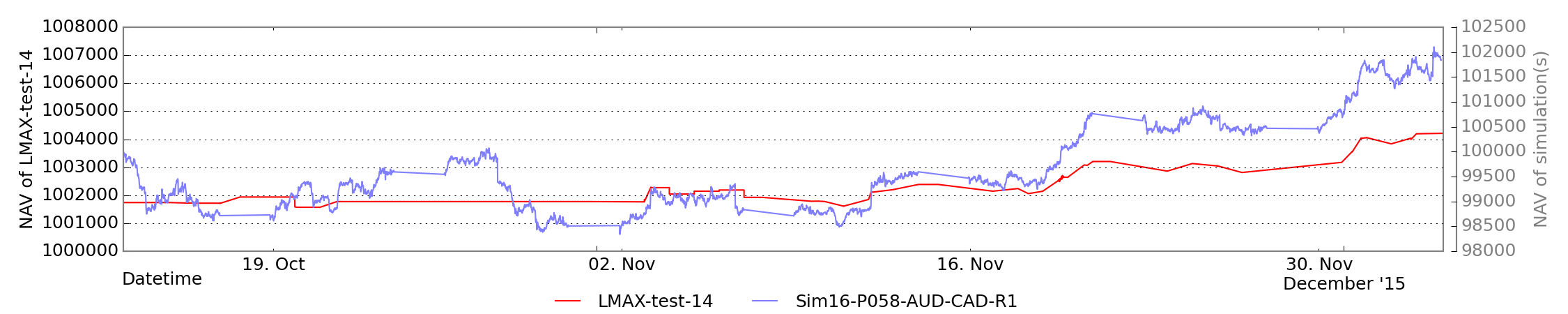}\\
	\includegraphics[width=1.5\columnwidth]{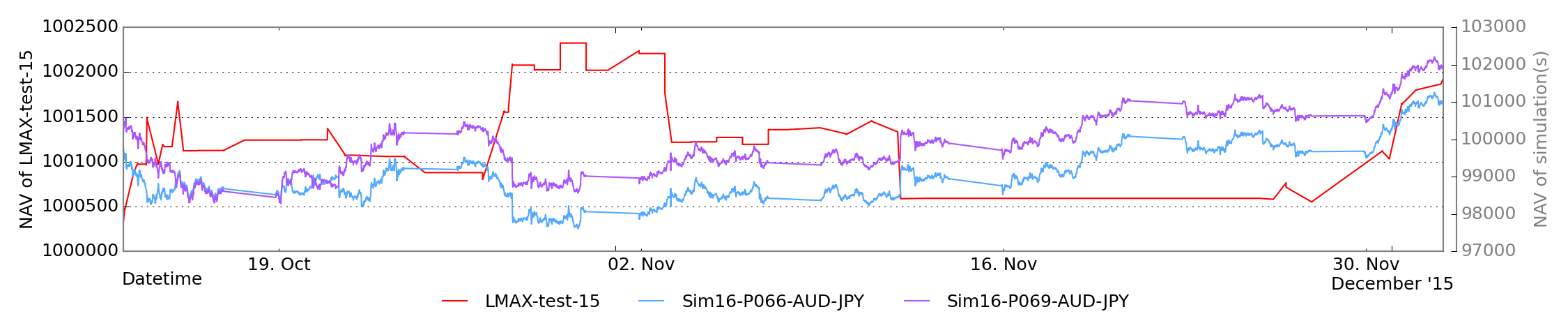}\\
	\includegraphics[width=1.5\columnwidth]{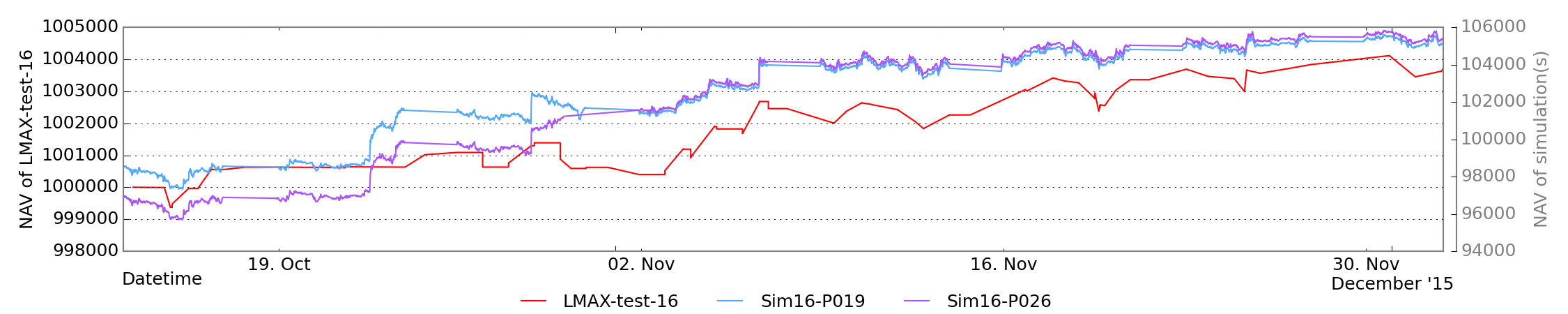}
	\caption{Net asset value plots for opened demo sessions and the results of simulations performed with StringAlgo v.16 on real demo trading IB and LMAX accounts as presented in Tab.~\ref{tab:demo1}.}\label{fig:demo1}
\end{figure*}

Fig.~\ref{fig:demo1} shows net asset value (NAV) plots for this real demo trading results and the simulation results (the NAV scales differ as they were initialized with different trade volumes). One can observe that all demo results for currency pairs follow the main trend of simulations for chosen time period, i.~e., nearly two months. The best coincidence is clearly visible for currency pair EUR/USD (IB-test-12 and LMAX-test-16 accounts), which was in the center of our interest. Also we found nice candidate for currency pair AUD-CAD as one can see in the case of LMAX-test-14 account.

The encouraging results with evolutionary algorithm for the parametric optimization \cite{Bundzel:2015bc} lead us to enhance the algorithm with a module for parallel evaluation of string moment values on short time scales. The trends in price change, identified either with volatility or angular momentum (Fig.~\ref{fig:regge1}), yield to dynamic change of the parameters as a string length and a trade altitude. They are not keep constant, e.~g., the trade altitude is lowered, so the algorithm can profit even under new conditions. The evolutionary algorithm has predefined limits within which it selects the most suitable combination of parameters leading ultimately to buy/sell orders. Intensive tests of new module are ongoing and we expect the first results in the near future.

\begin{table*}[thbp]
	\centering
	\begin{tabular}{r c c c c}
		\toprule
		\textbf{\ Demo session\ } & \textbf{\ Currency pair\ } & \textbf{\ Start of session\ } & \textbf{\ End of session\ } & \textbf{\ Simulation\ }\\
		\midrule
		IB-test-12            & EUR/USD         & 2015-09-28 	& 2015-11-23  & SIM16-P009 \\
		LMAX-test-16          & EUR/USD         & 2015-10-13 	& 2015-12-03  &	SIM16-P010 \\
		& & & &	SIM16-P019 \\
		& & & &	SIM16-P026 \\
		\midrule
		LMAX-test-13          & CHF/JPY         & 2015-10-09 	& 2015-12-01  & SIM16-P046-CHF-JPY \\
		& & & &	SIM16-P047-CHF-JPY \\
		& & & &	SIM16-P052-CHF-JPY \\
		\midrule
		LMAX-test-14          & AUD/CAD         & 2015-10-12 	& 2015-12-05 & SIM16-P058-AUD-CAD-R1\\
		\midrule
		LMAX-test-15          & AUD/JPY         & 2015-10-12 	& 2015-12-03 & SIM16-P066-AUD-JPY \\
		& & & & SIM16-P069-AUD-JPY \\
		\bottomrule
	\end{tabular}
	\caption{Summary of opened sessions for real demo trading on the Interactive Brokers and LMAX Exchange market accounts.}\label{tab:demo1}
\end{table*}

\begin{acknowledgments}
The work was partially supported by Slovak Research and Development Agency SRDA and Slovak Grant Agency for Science VEGA under the grants No. APVV-0463-12, VEGA 1/0158/13 and VEGA 2/0009/16. The authors thank the TH division for warm hospitality during their visits at CERN.
\end{acknowledgments}

%
%

\bibliographystyle{apsrev} 
\bibliography{timeSeries}

\end{document}